\documentclass{aa}  

\usepackage{graphicx}
\usepackage{txfonts}
\usepackage{comment}
\usepackage{lscape} 
\usepackage[english]{babel}
\usepackage{tabularx}
\usepackage{booktabs}
\usepackage{caption}
\usepackage{soul}
\usepackage{threeparttable}
\usepackage{multirow,multicol} 
\usepackage[dvipsnames]{xcolor}
\usepackage{hyperref}
\hypersetup{
    colorlinks=true,
    linkcolor=blue,
    filecolor=magenta,      
    urlcolor=cyan,
}

\begin{document} 

   \title{Broadband X-ray spectral variability \\ of the pulsing ULX NGC 1313 X-2}


   \author{A. Robba\thanks{alessandra.robba@inaf.it}  
          \inst{1,2}
          \and
          C. Pinto
          \inst{2}
          \and
          D. J. Walton
          \inst{3}          
          \and
          R. Soria
          \inst{4,5}
          \and
          P. Kosec
          \inst{3,13} 
          \and
          F. Pintore
          \inst{2}
          \and
          T. P. Roberts
          \inst{6}
          \and
          W. N. Alston
          \inst{3} 
          \and
          M. Middleton
          \inst{7}
          \and
          G. Cusumano
          \inst{2}
          \and
          H. P. Earnshaw
          \inst{8}
          \and
          F. F{\"u}rst
          \inst{9}
          \and
           R. Sathyaprakash
           \inst{6,12}
          \and
           E. Kyritsis
          \inst{10,11}
          \and
           A. C. Fabian
          \inst{3}
          }

   \institute{Universit\`a degli Studi di Palermo, Dipartimento di 
              Fisica e Chimica, via Archirafi 36, I-90123 Palermo, Italy
         \and
             INAF/IASF Palermo, via Ugo La Malfa 153, I-90146 Palermo, Italy
         \and
             Institute of Astronomy, Madingley Road, CB3 0HA Cambridge, UK
         \and
             College of Astronomy and Space Sciences, University of the Chinese Academy of Sciences, Beijing 100049, China
         \and
             Sydney Institute for Astronomy, School of Physics A28, The University of Sydney, Sydney, NSW 2006, Australia
         \and
                 Centre for Extragalactic Astronomy, Department of Physics, Durham University,South Road, Durham DH1 3LE, UK
         \and
                 Department of Physics and Astronomy, University of Southampton, Highfield, Southampton SO17 1BJ, UK
         \and
                 Cahill Center for Astronomy and Astrophysics, California Institute of Technology, Pasadena, CA 91125, USA
         \and
             Science Operations Department,European Space Astronomy Centre (ESA/ESAC), Villanueva de la Canada, 28692, Madrid, Spain
         \and
                 University of Crete, Department of Physics, GR-71003, Heraklion, Greece
         \and
                 Institute of Astrophysics, FORTH, GR-71110 Heraklion, Greece
             \and
                Institut de Ciències de l'Espai , Carrer de Can Magrans, 08193 Cerdanyola del Vallès, Barcelona
             \and
                MIT Kavli Institute for Astrophysics and Space Research, Cambridge, MA 02139, USA
             }

   \date{Received ?; Accepted ?}
   
   \abstract
{It is thought that ultraluminous X-ray sources (ULXs) are mainly powered by super-Eddington accreting neutron stars or black holes as shown by the recent discovery of X-ray pulsations and relativistic winds.}
{This work presents a follow-up study of the spectral evolution over two decades of the pulsing ULX NGC 1313 X-2 in order to understand the structure of the accretion disc. The primary objective is to determine the shape and nature of the dominant spectral components by investigating their variability with the changes in the source luminosity.}
{We performed a spectral analysis over the canonical 0.3--10.0 keV energy band of all the high signal-to-noise {\it XMM-Newton} observations ($96$\,\% of the available data), and we tested a number of different spectral models, which should approximate super-Eddington accretion discs. The baseline model consists of two thermal blackbody components with different temperatures plus an exponential cutoff powerlaw.}
{The baseline model provides a good description of the X-ray spectra. In particular, the hotter and brighter ($L_X \sim 6$--$9 \times 10^{39}$\,erg s$^{-1}$) thermal component describes the emission from the super-Eddington inner disc and the cutoff powerlaw describes the contribution from the accretion column of the neutron star. Instead, the cooler component describes the emission from the outer region of the disc close to the spherisation radius and the wind. 
The luminosity-temperature relation for the cool component follows a negative trend, which is not consistent with L $\propto$ T$^4$, as is expected from a sub-Eddington thin disc of Shakura-Sunayev. This is not consistent with L $\propto$ T$^2$ either, as is expected for an advection-dominated disc.\ However, this would rather agree with a wind-dominated X-ray emitting region. 
Instead, the (L$_x$,T$_{disk}$) relation for the hotter component is somewhere in between the first two theoretical scenarios. }
{Our findings agree with the super-Eddington scenario and provide further detail on the disc structure. The source spectral evolution is qualitatively similar to that seen in NGC 1313 X-1 and Holmberg IX X-1, indicating a common structure and evolution among archetypal ULXs.}   

   \keywords{Accretion, accretion discs -- X-rays: binaries -- X-rays: individual (NGC 1313 X-2)}

   \maketitle
%

\section{Introduction}
Ultraluminous X-ray sources (ULXs) are among the best candidates for studying super-Eddington accretion in stellar-mass accreting compact objects. 
ULXs are the brightest off-nuclear, steady, point-like X-ray sources ($>$ 10$^{39}$ erg s$^{-1}$) in the Universe. They are often found in or near regions of recent star formation \citep{Swartz_2009, Kovlakas_2020} and they have X-ray luminosities that exceed the isotropic Eddington luminosity for a standard black hole (BH) with a mass of M$\approx$10 M$_{\odot}$ (e.g. \citealt{Kaaret_2017}). 
ULXs represent a heterogeneous sample of astronomical sources and are composed of a compact object, most likely a BH or a NS, and a companion star, which has been found to be a red or blue supergiant in some ULXs, see for example \cite{Heida_2019}.

In order to explain the high X-ray luminosities of these sources, several hypotheses have been proposed. 
A first scenario suggests that ULXs are powered by stellar-mass BHs whose radiation is preferentially beamed in our line of sight \citep{King_2001, Poutanen_2007}. 
A second scenario supposes that a BH more massive than 10 M$_{\odot}$ (30$-$80 M$_{\odot}$; e.g. \citealt{Zampieri_2009}), which accretes at or below the Eddington limit, is the compact object of a ULX. The existence of more massive BHs was proven by the detection of gravitational waves (e.g. \citealt{Abbott_2016a, Abbott_2016b, Ligo_2018}, with BH masses between 10 M$_{\odot}$ and 80 M$_{\odot}$. In addition, in the past, other theories have suggested that some of these systems could be intermediate-mass BHs ($10^{3-4}$\,M$_{\odot}$, \citealt{Colbert_1999}), accreting at sub-Eddington rates from low-mass companion stars with ESO243-49 HLX-1 being the best candidate (see \citealt{Farrell_2009}). 

For many years, the mass estimation of putative BHs powering ULXs was the subject of significant debate. However, the recent detection of coherent pulsations in several sources clearly demonstrates that some ULXs are powered by NSs accreting at very high Eddington rates with luminosities up to $\sim$\,500\,L$_{\rm Edd}$. 
The first pulsation in a ULX was discovered in M\,82 X-2 by \cite{Bachetti_2014} with {\it NuSTAR} observations and, at the moment, six ULXs are known to exhibit pulsations \citep{Israel_2016a, Israel_2016b, Furst_2016, Carpano_2018, Sathyaprakash_2019}. 
It is not straightforward as to how to distinguish between BH and NS accretors based on the spectral analysis alone \citep{Pintore_2017,Koliopanos_2017,Walton_2018b}. Moreover, pulsations are not always detectable; high count rates are indeed required for a low pulsed fraction or long exposure times for low count rates. This means that NSs are likely numerous among the compact objects of ULXs (see also \citealp{Rodriguez_2020,Wiktorowicz_2019,King_2017,Middleton_2016}).

One of the fundamental predictions of the super-Eddington accretion theory is that strong, relativistic winds are launched from the supercritical discs, driven by the extreme radiation pressure (see e.g. \citealp{Poutanen_2007}). \cite{Middleton_2014} suggest that the spectral residuals around 1 keV could be associated with the winds. The first discovery of powerful winds in two ULXs was achieved by \cite{Pinto_2016} by detecting blueshifted absorption lines in the high-resolution soft X-ray spectra provided by the {\it XMM-Newton} Reflection Grating Spectrometers (RGS). Further confirmations in other ULXs and with different detectors were obtained by \cite{Walton_2016}, \cite{Pinto_2017}, \cite{Kosec_2018a} and \cite{Kosec_2018b}. The exact launching mechanism of such winds is still unclear as magnetic pressure might also contribute, although \citet{Pinto_2020} show that the relation between their velocities and ionisation parameters with the ULX luminosities agrees with the radiation driving mechanism. A thorough understanding of the ULX phenomenology requires additional constraints on the nature of these winds and their link with the source appearance and, therefore, accretion rate. This involves a study of the wind properties via high-resolution X-ray spectroscopy combined with a careful study of the evolution of ULX broadband spectra. 

\begin{figure}
   \centering
   \includegraphics[width=8.5cm]{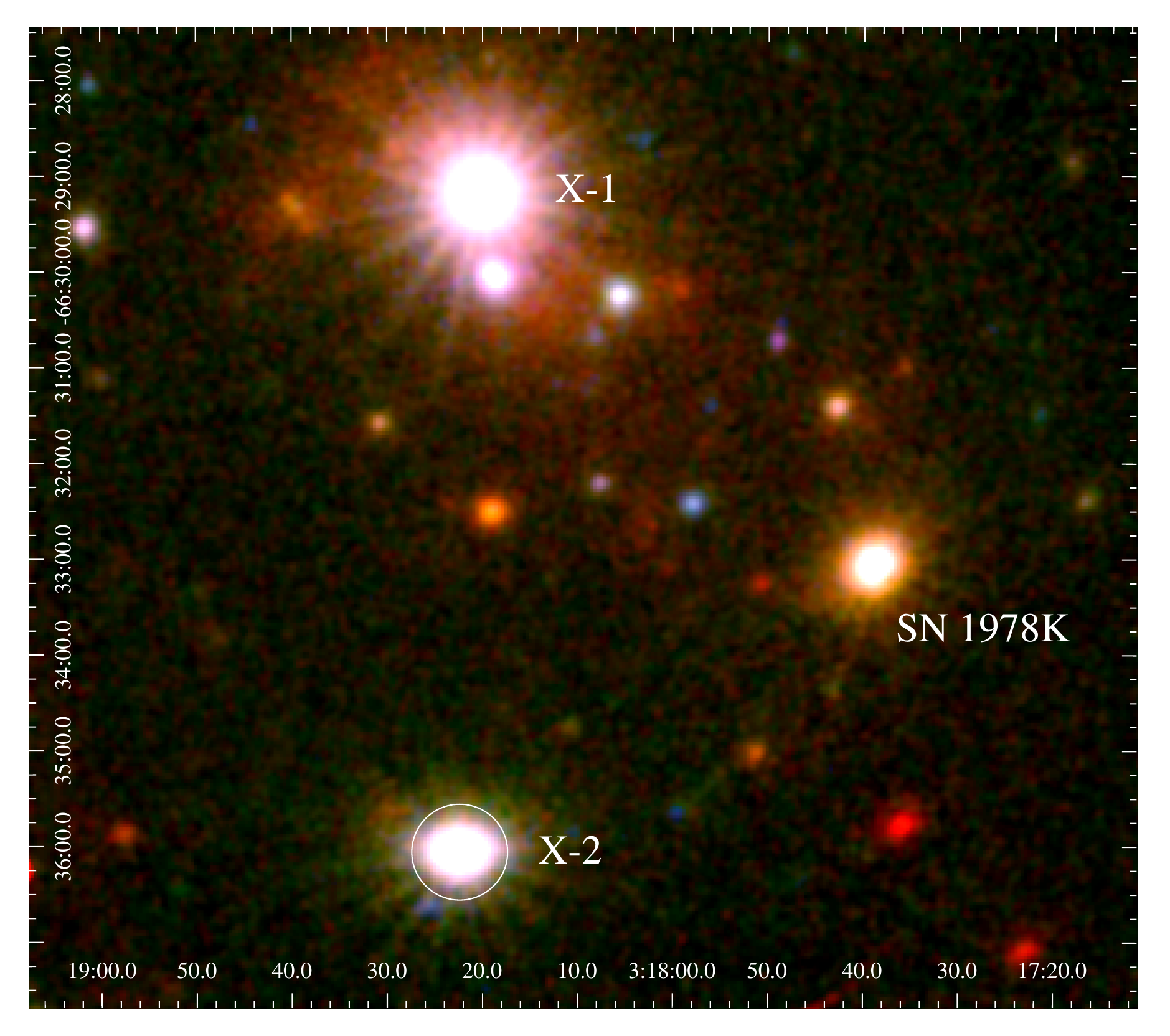}
      \caption{{\it XMM-Newton} image of NGC 1313, which we obtained by combining all the data available from the 2017 EPIC-pn and MOS 1,2 observations. The 30$^{\prime\prime}$-radius circle around the ULX X-2 represents our default source extraction region. The red colour corresponds to 0.2--1 keV, green is for 1--2 keV, and blue is for 2--12 keV.}
    \label{Campo1}
   \end{figure}

In this work we present the analysis of the X-ray spectra of the pulsating ULX X-2 in the galaxy NGC 1313. This barred galaxy (see Fig. \ref{Campo1}) hosts a supernova remnant (SN  1978K) and two ULXs: X-1, close to the nucleus, and X-2, in the outskirts of the galaxy, which is the subject of this work.  \cite{Sathyaprakash_2019} discovered pulsations in NGC 1313 X-2 for the first time thanks to our deep {\it XMM-Newton} campaign. This ULX is characterised by strong variability in both luminosity and spectral shape. The high X-ray variability and spectral hardness suggest that the object is viewed at an inclination angle, which is low enough to allow for a direct view of the inner regions of its accretion flow and to detect pulsations \citep{Middleton_2015b}. NGC 1313 X-2 also shows evidence of winds in the form of X-ray spectral features in the soft band (see, e.g. \citealt{Middleton_2015b} and \citealt{Kosec_2018a}).

Throughout this work, we assume a distance of D = 4.2 Mpc to NGC 1313 \citep{Mendez_2002, Tully_2016}. The only Cepheid distance available is 4.6 Mpc, which is a little bit of an outlier from the other measurements \citep{Qing_2015}.

This paper is structured as follows. In Section \ref{DATA ANALYSIS}, we provide details about the {\it XMM-Newton} observations used in this work and the data reduction. We show some basic time properties and the model-independent variability of the X-ray spectral shape. In Section \ref{MAIN RESULTS}, we describe the main results of the spectral and timing analysis, while in Section \ref{DISCUSSION} we discuss the behaviour of the thermal components in the luminosity-temperature plane and, finally, provide our conclusions in Section \ref{CONCLUSIONS}.

\section{Data analysis}\label{DATA ANALYSIS}
\subsection{Observations}

We analysed the public archival {\it XMM-Newton} data of all the high signal-to-noise  observations between 2000 and 2017. For our analysis, we particularly benefitted from a recent deep $\sim$1Ms view of NGC 1313 (PI: Pinto). 
The observations were carried out with the EPIC-pn and EPIC-MOS detectors \citep{Struder_2001,Turner_2001}.
For observations 0803990301 and 0803990401, we used only the data provided by pn and MOS 2 since the source was out of the MOS1 field of view due to damage to CCD6 and CCD3 (since 2005 and 2012, respectively). 
We have excluded observations 0150280201 and 0150280701 from this analysis as they contain only MOS data and they were very short, therefore providing low statistics.\ We also excluded observation 0205230201 because of a slew failure. Table \ref{XMM_observations} lists the details of the high signal-to-noise {\it XMM-Newton} observations ($96$\,\% of the available data) that we analysed, including the date, the duration after the removal of flares, the count rates, and the filter of each observation.  

\begin{table*}
\centering
\caption{\emph{XMM-Newton} observations of NGC 1313 X-2.}
    \begin{tabular}{ccccccccc}
    \toprule
    Obs.ID${(1)\atop}$ & Date${(2)\atop}$ & \multicolumn{3}{|c|}{Exposure time${(3)\atop}$ (s)} & \multicolumn{3}{|c|}{Count rates${(4)\atop}$ (cts/s)} & Filter \\
    & & pn & MOS1 & MOS2 & pn & MOS1 & MOS2 & \\
    \midrule
    0106860101 & 2000-10-17 & 31638 & 29250 & 29183 & 0.2554 & 0.07654 & 0.07588 & Medium \\
    0150280101 & 2003-11-25 & 19173 & 12498 & 12501 & 0.712 & 0.2181 & 0.2584 & Thin1 \\
    0150280201 & 2003-12-09 & 5648  & 3609  & 3622  & -     & -      & -      & Thin1 \\
    0150280301 & 2003-12-21 & 10335 & 11971 & 11976 & 0.8365 & 0.2803 & 0.2795 & Thin1 \\
    0150280401 & 2003-12-23 & 14098 & 15278 & 15293 & 0.9459 & 0.3081 & 0.3104 & Thin1 \\
    0150280501 & 2003-12-25 & 15299 & 15262 & 15249 & 0.5249 & 0.1551 & 0.1510 & Thin1 \\
    0150280601 & 2004-01-08 & 14757 & 15382 & 15377 & 0.4095 & 0.1277 & 0.1287 & Thin1 \\
    0150280701 & 2003-12-27 & 16686 & 17779 & 17810 & -     & -      & -      & Thin1 \\
    0150281101 & 2004-01-16 & 7036 & 8671 & 8676 & 0.3545 & 0.1117 & 0.1170 & Thin1 \\
    0205230201 & 2004-05-01 & 3459 & 12270 & 12275 & -     & -      & -      & Thin1 \\
    0205230301 & 2004-06-05 & 10036 & 11672 & 11674 & 0.9813 & 0.3181 & 0.32 & Thin1 \\
    0205230401 & 2004-08-23 & 16137 & 17771 & 17776 & 0.2829 & 0.08781 & 0.09462 & Thin1 \\
    0205230501 & 2004-11-23 & 14137 & 15769 & 15774 & 0.3190 & 0.09469 & 0.09883 & Thin1 \\
    0205230601 & 2005-02-07 & 12437 & 14071 & 14074 & 0.9053 & 0.2974 & 0.3023 & Thin1 \\
    0301860101 & 2006-03-06 & 19937 & 21570 & 21575 & 0.6510 & 0.1734 & 0.1963 & Medium \\
    0405090101 & 2006-10-15 & 121190 & 122456 & 122453 & 0.6437 & 0.1689 & 0.1576 & Medium \\
    0693850501 & 2012-12-16 & 123341 & 124921 & 124927 & 0.5395 & 0.1820 & 0.1819 & Medium \\
    0693851201 & 2012-12-22 & 123341 & 124921 & 124924 & 0.275 & 0.1036 & 0.1086 & Medium \\
    0722650101 & 2013-06-08 & 28841 & 30421 &30426  & 0.1713 & 0.009892 & 0.01064 & Medium \\
    0742590301 & 2014-07-05 & 60040 & 61653 & 61624 & 0.3937 & 0.1074 & 0.1035 & Medium \\
    0742490101 & 2015-03-30 & 100041 & 98937 & 101625 & 0.07816 & 0.07223 & 0.07989 & Medium \\
    0764770101 & 2015-12-05 & 71941 & 73555 & 73524 & 0.2417 & 0.07106 & 0.07295 & Thin1 \\
    0764770401 & 2016-03-23 & 30041 & 31653 & 31624 & 0.4102 & 0.1260 & 0.1261 & Thin1 \\
    0782310101 & 2016-10-08 & 88041 & 89655 & 89626 & 0.5705 & 0.1863 & 0.1865 & Medium \\
    0794580601 & 2017-03-29 & 44542 & 46155 & 46127 & 0.3424 & 0.07281 & 0.1022 & Medium \\
    0803990101 & 2017-06-14 & 134142 & 133036 & 135725 & 0.2304 & 0.09635 & 0.09032 & Medium \\
    0803990201 & 2017-06-20 & 130841 & 132453 & 132425 & 0.3525 & 0.07631 & 0.08879 & Medium \\
    0803990301 & 2017-08-31 & 96686 & - & 91971 & 0.2380 & - & 0.07478 & Medium \\
    0803990401 & 2017-09-02 & 64008 & - & 63472 & 0.1077 & - & 0.09849 & Medium \\
    0803990701 & 2017-09-24 & 14500 & 9798 & 9797 & 0.2483 & - & 0.06602 & Medium \\
    0803990501 & 2017-12-07 & 125941 & 127555 & 127524 & 0.07091 & 0.07314 & 0.07887 & Medium \\
    0803990601 & 2017-12-09 & 125942 & 127555 & 127524 & 0.07893 & 0.07587 & 0.08123 & Medium \\
    \bottomrule  
    \end{tabular}
    \caption* {\footnotesize {Notes: ${(1)\atop}$ observation identifier; ${(2)\atop}$ observation date (yyyy-mm-dd), ${(3)\atop}$ exposure time corrected for solar flares, ${(4)\atop}$ count rates in the  0.3--10 keV energy band}.}
    \label{XMM_observations}
\end{table*}

\subsection{Data reduction}\label{DATA REDUCTION}

The data analysis was performed using the \textit{XMM-Newton Science Analysis System} (SAS) version 18.0.0 and the calibration files of January 2020.\footnote{https://www.cosmos.esa.int/web/xmm-newton/ccf-release-notes.} Following the standard procedures, EPIC-pn and MOS data were reprocessed using the tasks \textit{'epproc}' and '\textit{emproc}'. The calibrated and concatenated event lists were filtered for high background epochs to acquire good time intervals (GTI) as follows. For each data set and each instrument, we extracted the high energy light curve (including events between 10-12 keV) to identify intervals of flaring particle background and we chose a suitable threshold (0.35 and 0.40 cts \textbf{s$^{-1}$} for EPIC-MOS and pn, respectively), which is above the low steady background, to create the corresponding filtered EPIC event list. As recommended, we selected only single and double events (PATTERN $\leq$4) for EPIC-pn, and single to quadruple events (PATTERN $\leq$12) for EPIC-MOS. 

We extracted EPIC MOS 1-2 and pn images in the 0.2--1/1--2/2--12\,keV energy range and stacked them with the '\textit{emosaic}' task. The field of the NGC 1313 galaxy and the brightest X-ray sources are shown in Fig.\,\ref{Campo1}.
We generally extracted source spectra from circular regions with a radius of $30"$, except where the source was near the edge of the CCD (in these cases we used regions with a radius of $20"$) and the corresponding background from a larger circle in a nearby region on the same chip, free from contaminating point sources. The background region was also not generally placed in the Copper emission region \citep{Lumb_2002}, with the exception of a small number of observations for which X-2 was also located in the Cu region. 

We used the task '\textit{arfgen}' to reproduce the effective area of the instrument and to correct instrumental factors, such as bad pixels and bad columns, using calibration information. 
The response matrix was generated with '\textit{rmfgen}'. Since the EPIC-MOS1 and EPIC-MOS2 spectra are consistent for each observation, we stacked data from the MOS cameras into a single spectrum with the '\textit{epicspeccombine}' routine.

\subsection{Spectral analysis}\label{Spectral analysis}

\begin{figure*}
   \centering
   \includegraphics[width=20cm]{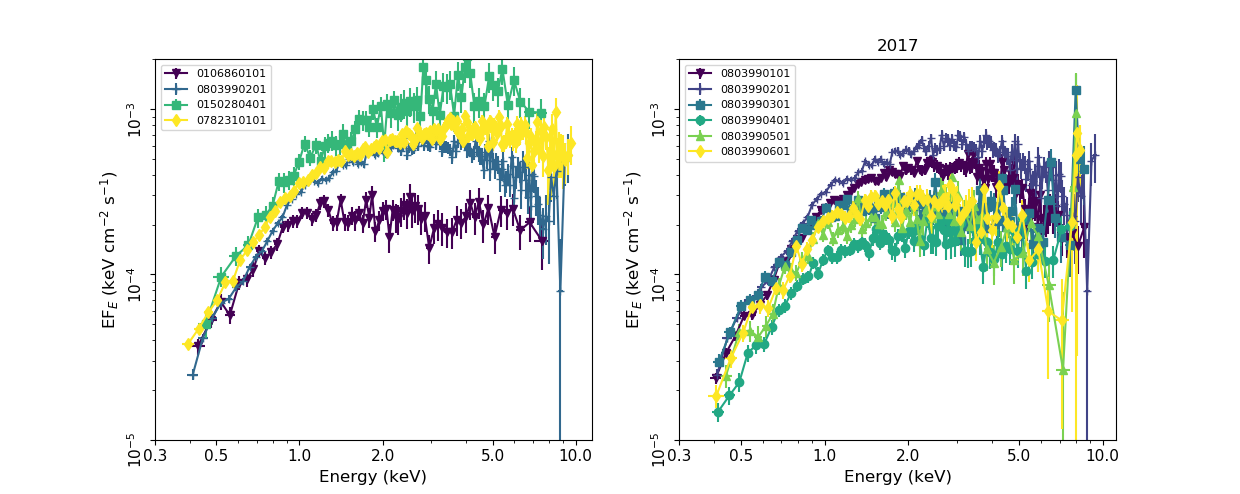}
      \caption{Left panel: Comparison between four spectra, from low (Obs.ID:0106860101), to intermediate (Obs.ID:0803990201 and Obs.ID:0782310101), and to high flux (Obs.ID:0150280401). Right panel: {\it XMM-Newton}/EPIC-pn spectra of the recent observations (2017 campaign).}
    \label{Confronto2}
   \end{figure*}

Fig. \ref{Confronto2} shows an overview of the spectral properties of X-2. In particular, the right panel illustrates the shape of the observed EPIC-pn spectra of NGC 1313 X-2 for six individual exposures during the most recent campaign (2017). 
The spectra indicate substantial variability in luminosity by a factor of up to five and there is no significant spectral variability. The left panel of Fig. \ref{Confronto2} instead shows a comparison between some spectra of a remarkably different shape and flux: a high flux spectrum (Obs.ID:0150280401), two spectra with intermediate flux (Obs.ID:0803990201 and Obs.ID:0782310101), and one with low flux (Obs.ID:0106860101). 
As it has been observed in most ULXs (e.g. \citealt{Middleton_2015b}), the spectrum becomes harder at higher fluxes. This behaviour disagrees with that seen in the classical Galactic X-ray binaries that accrete below Eddington limit and it is therefore considered as strong evidence in support of super-Eddington accretion.
The left panel of Fig. \ref{Confronto2} clearly shows that the spectrum is more variable in the hard X-ray band ($\gtrsim$ 1 keV), which indicates that at least two different (soft and hard) spectral components are responsible for the X-ray emission.

We carried the spectral analysis out over the 0.3-10.0 keV energy range with {\sc XSPEC} version 12.10.1 (\citealt{Arnaud_1996}). We simultaneously fitted the spectra of the MOS and pn cameras and estimated the parameter uncertainties at the 68$\%$ confidence level. 
Spectra were grouped for a minimum of 25 counts per energy bin, so that the $\chi^2$ statistic could be used.

There is no coverage beyond 10 keV in most observations and even in the {\it NuSTAR} data, the source has low statistics above 10 keV (e.g. \citealt{Bachetti_2013}). As the source is generally softer than X-1 in the NGC 1313 galaxy, the systematic error due to the lack of {\it NuSTAR} data is negligible.
 We therefore focussed on the canonical 0.3-10 keV X-ray band and assumed a simplified model with either two or three emission components.  In the case, for instance, of the double blackbody disc and an exponential cutoff powerlaw model, the low-temperature component mimics the emission of the outer disc and wind, while the hotter temperature component reproduces the emission from the inner super-Eddington accretion flow. The third component is necessary to describe the contributions of the central accretion columns by the magnetic accretor, that is, the NS. \cite{Walton_2018b} show that in all ULXs, this latter component is significant above 8-9 keV (see also \citealt{Bachetti_2013}).

\subsection{Timing analysis}\label{Timing analysis}

\begin{figure*}[ht!]
\centering
\includegraphics[width=18cm]{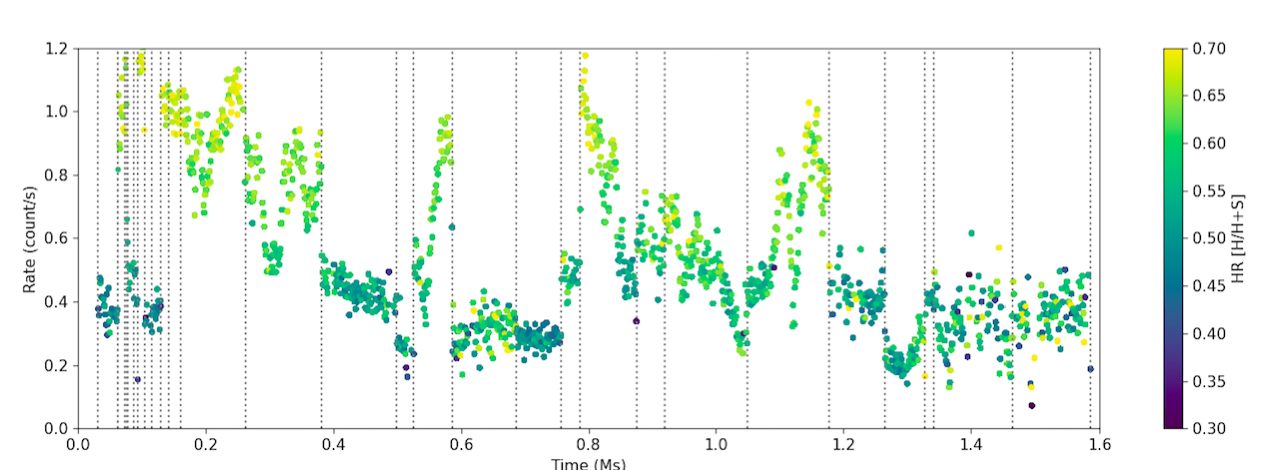}
\caption{0.3--10 keV long-term EPIC/pn light curve of NGC 1313 X-2 colour-coded according to the HR for all the {\it XMM-Newton} observations, with time bins of 1000 s. The HR was computed using the light curves in the soft [0.3--1.2 keV] and in the hard [1.2--10 keV] X-ray energy bands. Observations occur at different epochs: we removed the gaps between observations with grey-dashed lines for visual purposes.}
\label{lightcurve}
\end{figure*}

In order to better understand our spectra, we extracted light curves in the whole band (0.3--10 keV) and in two different bands (i.e. soft 0.3--1.2 keV and hard 1.2--10 keV). The latter were extracted to calculate the hardness ratios (HRs), which were computed as follows:
\begin{equation}
HR = \frac{Hard\,Rate_{\,1.2-10\,\mathrm{keV}}}{Soft\, Rate_{\,0.3-1.2\,\mathrm{keV}}+Hard\,Rate_{\,1.2-10\,\mathrm{keV}}}
.\end{equation}
We chose to split the bands at 1.2 keV because this is the average energy at which the spectral curvature changes in the EPIC spectra (the case for Obs.ID:0803990201 is shown in Fig. \ref{grafico_confronto}). We plotted the full-band light curve colour-coded according to the HR in Fig. \ref{lightcurve}, where the vertical grey-dashed lines separate  each observation.

NGC 1313 X-2 shows high variability during the two decades of (non-contiguous) observations. As a short-term flux variability test, we adopted the normalised excess variance in the whole band (0.3--10 keV), which is useful to quantify the different amplitudes of intrinsic variability of each light curve. As reported by \cite{Nandra_1997}, it is defined as follows:
\begin{equation}
        \sigma^2_{NXV} = \frac{1}{N\bar{x}^2} \sum_{i=1}^{N} [(x_i - \bar{x})^2 - \sigma^2_{{err},i}]
        \label{}
,\end{equation}

where x$_i$ and $\sigma_{{err},i}$ are the count rate and its error in the i-th bin, $\bar{x}$ is the mean count rate, and N is the number of bins used to estimate $\sigma _{NXV}^2$. The associated error is the following:
\begin{gather*}
\Delta\sigma _{NXV}^2 = \frac{S_D}{\bar{x}^2(N)^{1/2}}, \\
S_D = \frac{1}{N-1} \sum_{i=1}^{N}\{[(x_i - \bar{x})^2 - \sigma^2_{{err},i}] - \sigma _{NXV}^2\bar{x}^2\} ^2.
\label{}
\end{gather*}

The fractional root mean square (RMS) variability amplitude ($F_{var}$, see \citealt{Vaughan_2003}) is the square root of the normalised excess variance, that is
\begin{equation}
        F_{var} (\%) = \sqrt{\sigma^2_{NXV}} \times 100 = \sqrt{\frac{1}{N\bar{x}^2} \sum_{i=1}^{N} [(x_i - \bar{x})^2 - \sigma^2_{{err},i}]} \: \times 100
        \label{eq_Fvar}
\end{equation}
\begin{equation}
        \Delta\sigma_{F_{var}} (\%) = \frac{1}{2}*\frac{\Delta\sigma _{NXV}^2}{\sqrt{\sigma^2_{NXV}}} \: \times 100
        \label{}
,\end{equation}

where F$_{var}$ is a linear statistic, which gives the same information as $\sigma^2_{NXV}$, but in percentage terms. In order to compare the RMS and to account for the different exposure times, we split all light curves into 40ks segments, which is  approximately a minimum-common-denominator segment for the long observations.\ Then we averaged the RMS results from the two or more segments available in longer (80-120 ks) observations and binned with $\Delta T$  of 1000 s. For the observations shorter than 40 ks, we did not estimate the RMS.

\section{Main results}\label{MAIN RESULTS}

\subsection{Spectral modelling}
In this section we present the spectral analysis of NGC 1313 X-2. In order to obtain an adequate description of the continuum, ULX broadband X-ray spectra require several emission components. Among the several models tested, we adopted one similar model to that used in \cite{Walton_2018b}. All models also include neutral absorption, which was modelled with TBABS \citep{Wilms_2000}. This absorption, which is due to an interstellar and circumstellar medium in the line of sight towards the source, is necessary to partially explain the low-energy spectral curvature. We adopted a lower limit equal to the Galactic value of N$_H$=7 $\times$ 10$^{20}$ cm$^{-2}$.

\subsubsection{Baseline models}\label{Baseline_models}

Our baseline model for the X-ray spectra of NGC 1313 X-2 consists of two thermal components, a cold disc blackbody (DISKBB) and a warmer DISKBB for the outer and the inner disc, respectively.\ It also consists of an exponential cutoff powerlaw component (CUTOFFPL) for the central accretion column that forms when the material flows down onto the magnetic poles. For the CUTOFFPL component, we set its spectral parameters to the average values seen from the pulsed emission of the following ULX pulsars currently known: $\Gamma$ = 0.59 and E$_{cut}$= 7.9 keV (\citealt{Brightman_2016, Walton_2018a, Walton_2018b, Walton_2018c}).

In most spectra, there are notable residuals at $\sim$1 keV, related to atomic emission and absorption associated with the likely presence of an extreme outflow powered by wind \citep{Pinto_2016, Kosec_2018a, Kosec_2018b, Wang_2019}. Despite this, the overall shape of the X-ray spectra is well reproduced by our featureless continuum model. Indeed, as pointed by \cite{Pinto_2020} and \cite{Walton_2020}, the presence of the wind has no dramatic effects on the modelling of the spectral continuum and, therefore, we did not include any line emission or absorption components in our current fits. In addition, this would require the use of deep, on-axis, RGS spectra that are unavailable since the source is off-axis in most of the observations. This is valid for the optically thin ionised wind phase responsible for the spectral lines, but it might not hold in the case of further optically thick wind components.

The spectral fits with the best-fitting baseline double disc-blackbody and the exponential cutoff powerlaw models are presented in Appendix \ref{appendix_bestfit}. As we show in Table \ref{table_bestfit}, we obtained a column density between N$_H$ = (0.095--0.32) $\times$ 10$^{22}$ cm$^{-2}$, a cold temperature in the range from 0.20 < T$_1$ < 0.49 keV, and a warm temperature in the 0.69 < T$_2$ < 1.9 keV range. The corresponding luminosities are L$_1$ = (1--6) $\times$ 10$^{39}$ erg s$^{-1}$ and L$_2$ = (2--8) $\times$ 10$^{39}$ erg s$^{-1}$. The goodness of the spectral fits is indicated by the $\chi^2$, which are satisfactory as shown in Table \ref{table_bestfit} (the reduced $\chi^2$ are in the range from 0.8-1.4).

\subsubsection{Alternative models}\label{Alternative models}
\begin{figure*}[ht!]
        \centering
        {\includegraphics[width=.51\textwidth]{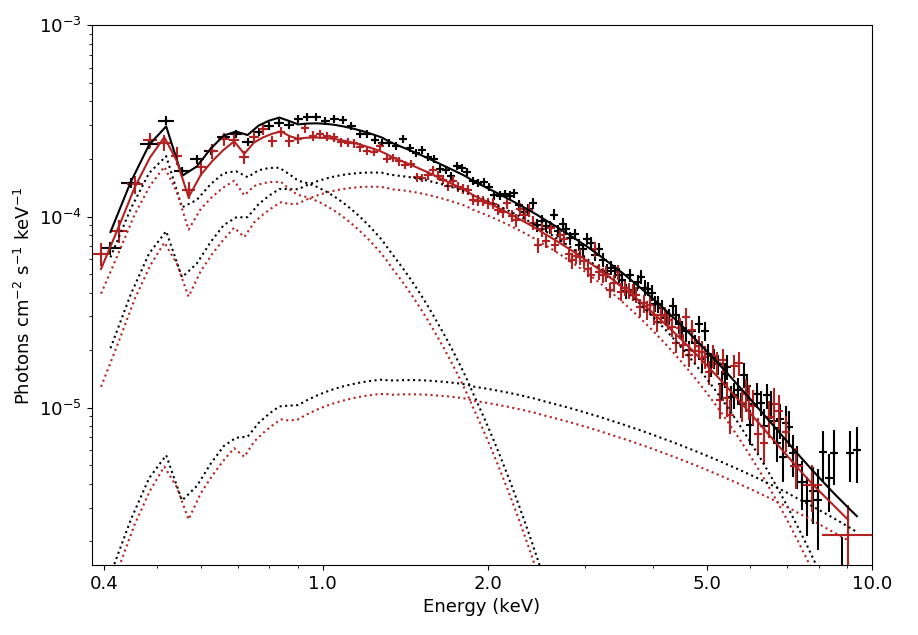}} 
        {\includegraphics[width=.48\textwidth]{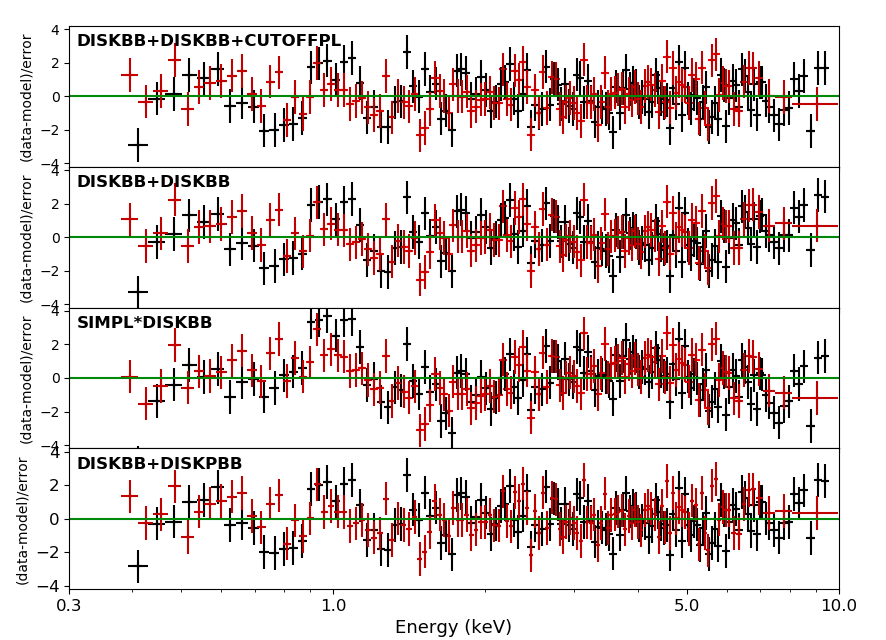}} 
    \caption{Left panel: \textit{XMM–Newton} unfolded spectrum of observation 0803990201 of NGC 1313 X-2. The black points are EPIC-pn and the red points show the stacked EPIC-MOS1/2 data. We overlapped the best-fit DISKBB+DISKBB+CUTOFFPL model (black and red, solid lines) and its single components (dashed lines). Right panel: Residuals from a selection of models listed in Table \ref{spectralmodels_0803990201}.}
    \label{grafico_confronto}
\end{figure*}

We also tested several other models to describe the spectral shape of the spectrum of X-2.  We show examples of our model fits for Obs.ID:0803990201 in Fig. \ref{grafico_confronto} with corresponding best-fit parameters in Table \ref{spectralmodels_0803990201}.

Spectral modelling of ULXs using {\it XMM-Newton} data is well established with two-component models, which resemble multi-colour blackbody emission (e.g. \citealt{Gladstone_2009} and \citealt{Stobbart_2006}). In this context, the first attempt consists of a double disc-blackbody DISKBB+DISKBB, in which the first component describes the outer accretion flow and the possible optically thick wind and the second one takes both the inner flow and any other emission closer to the NS into account. As we expected (considering the lack of the hard cutoff powerlaw component), in comparison to the best-fitting model, the values of the temperatures are larger. The column densities, on the other hand, are similar (see Table \ref{table_2diskbb}). We also tested the model with the column density fixed to N$_H$ = 1.94 $\times$ 10$^{21}$ cm$^{-2}$, obtained from the weighted average of the previous fits with free N$_H$, to evaluate the contribution of neutral absorption and systematic effects. However, we note that this assumption does not strongly influence the broadband continuum fits, so we preferred to keep N$_H$ free to vary amongst the spectra.

The spectral fits with two thermal components are generally worse than the baseline three component model with the greatest discrepancy at the lowest flux, where the hard cutoff powerlaw component starts to be important ($\Delta \chi^2$ = 41 and 61 for Obs.ID 0405090101 and 0782310101, respectively, for 1 additional degree of freedom). Spectral parameters are reported in Appendix \ref{appendix_nHfree} (see Table \ref{table_2diskbb}).

The second alternative model consists of replacing the double DISKBB continuum model with a single DISKBB modified by a SIMPL component. This empirical model of Comptonisation assumes that a fraction of the soft photons is scattered into a corona or a photosphere warmer than the disc.
As shown in Table \ref{spectralmodels_0803990201}, although the parameters are physically acceptable, we found a worse fit with an increase of $\chi^2$ ($\Delta \chi^2$ = 140) in observation 0803990201. Column densities, N$_{H}$, and cool temperature values are similar and follow the same behaviour as the baseline model with two DISKBBs plus a CUTOFFPL component. 

Afterwards, we tried using a DISKBB+DISKPBB model, where DISKPBB is a multiple blackbody disc model characterised by the additional free parameter p \citep{Mineshige_1994}. The $p$ parameter defines the radial dependency of the temperature, following the law T $\propto$ R$^{-p}$ (p=0.75 for a standard disc, p<0.75 for a profile affected by advection, and p=0.5 for a slim disc, see \citealt{Abramowicz_1988}). 

In this case, the temperatures obtained for the cool component are similar to parameters from the best-fit (in the DISKBB+DISKBB+CUTOFFPL baseline model); whereas for the hot component, the temperature is largely unconstrained in some observations.
In general, the values for parameter p (e.g. p=0.573$\pm$0.019, for Obs.ID:0803990201) indicate a regime very close to the slim disc profile (the mean value for all spectral fits is p$_{mean}$ $\sim$ 0.57). The addition of the extra free parameter (p) provides spectral fits that are statistically halfway between the two DISKBB and three component models.  

All the models show some residuals around 1 keV, as described in Section \ref{Baseline_models}. They appear very similar and narrow ($\sim0.1$\,keV) in all fits, indicating that they are independent from the particular model chosen and do not affect our results.


\begin{table}[h!]
\centering
\caption{Best-fit parameters for the models tested in this work for Obs.ID:0803990201.}
    \resizebox{\columnwidth}{!}{
    \begin{tabular}{cccc}
    \toprule
    \toprule
     \multicolumn{4}{c}{TBABS*(DISKBB+DISKBB+CUTOFFPL)} \\
    \midrule
    Model component & Parameter & Unit & \\
    \midrule
    TBABS  & N$_{H}$  & [10$^{22}$cm$^{-2}$] & 0.28 $\pm$ 0.02 \\
    DISKBB & T$_{in}$ & [keV]                & 0.25 $\pm$ 0.02 \\
           & norm     &                      & 14${+7\atop-5}$ \\
    DISKBB & T$_{in}$ & [keV]                & 1.13 $\pm$ 0.04 \\
           & norm     &                      & 0.053 $\pm$ 0.007 \\ 
    CUTOFFPL & PhoIndex &                    & 0.59 (fixed) \\
             & HighECut & [keV]              & 7.9 (fixed)  \\ 
             & norm   & [10$^{-5}$]          & 2.8${+0.4\atop-0.5}$ \\
    \midrule
    $\chi^2$/dof & & & 314.91/259\\
    \toprule
    \toprule   
    \multicolumn{4}{c}{TBABS*(DISKBB+DISKBB)}\\
    \midrule
    Model component & Parameter & Unit & \\
    \midrule
    TBABS  & N$_{H}$  & [10$^{22}$cm$^{-2}$] & 0.242${+0.014\atop-0.013}$ \\
    DISKBB & T$_{in}$ & [keV]                & 0.32$\pm$0.02 \\
           & norm     &                      & 4.3${+1.7\atop-1.2}$ \\
    DISKBB & T$_{in}$ & [keV]                & 1.36${+0.03\atop-0.02}$ \\
           & norm     &                      & 0.028 $\pm$ 0.002 \\ 
    \midrule
    $\chi^2$/dof & & & 336.86/260\\
    \toprule
    \toprule
    \multicolumn{4}{c}{TBABS*(SIMPL*DISKBB)}\\
    \midrule
    Model component & Parameter & Unit & \\
    \midrule
    TBABS & N$_{H}$   & [10$^{22}$cm$^{-2}$] & 0.158${+0.002\atop-0.004}$ \\
    SIMPL & $\Gamma$  &                      & 2.75${+0.07\atop-0.02}$ \\
    DISKBB & T$_{in}$ & [keV]                & 0.56${+0.01\atop-0.02}$ \\
           & norm     &                      & 0.62${+0.09\atop-0.08}$ \\ 
    \midrule       
    $\chi^2$/dof & & & 455.69/260 \\
    \toprule
    \toprule
    \multicolumn{4}{c}{TBABS*(DISKBB+DISKPBB)}\\
    \midrule
    Model component & Parameter & Unit & \\
    \midrule
    TBABS   & N$_{H}$  & [10$^{22}$cm$^{-2}$] & 0.32${+0.03\atop-0.02}$ \\
    DISKBB  & T$_{in}$ & [keV]                & 0.21 $\pm$ 0.02 \\
            & norm     &                      & 30${+37\atop-15}$ \\
    DISKPBB & T$_{in}$ & [keV]                & 1.57${+0.07\atop-0.05}$ \\
            & p        &                      & 0.57 $\pm$ 0.02 \\   
            & norm     &                      & 0.008  $\pm$ 0.002 \\
    \midrule
    $\chi^2$/dof & & & 321.32/259\\
    \bottomrule
    \end{tabular}}
    \label{spectralmodels_0803990201}
\end{table}

\subsection{Time variability}

NGC 1313 X-2 shows high variability during the two decades of (non-contiguous) observations, as seen in the long-term light curve (see Fig. \ref{lightcurve}). Strong short-term ($\sim$ hours) variability is also seen during some individual and long observations. 
We see that the source becomes harder when it is brighter, which is indicative of an increasingly brighter super-Eddington inner disc.  This is because the hot disc is the dominant component in the 0.3--10 keV energy band. 

The timing properties of X-2 were also probed using the fractional variability (see Section \ref{Timing analysis}). The average values of F$_{var}$ for several observations and the corresponding HR mean are presented in Table \ref{Fvar_mean}. We have reported only the values for the observations that have a common time baseline of exposure time (40 ks) for comparing the RMS estimated. The complete values calculated for each segment are shown in Appendix \ref{appendix_Fvar} (see Table \ref{Fvar}).
Also, in cases where the observed variance is less than the error associated with each bin, the excess variance value is negative. These values are excluded from our considerations. 

\begin{figure}[h!]
\centering
    \includegraphics[width=9.5cm]{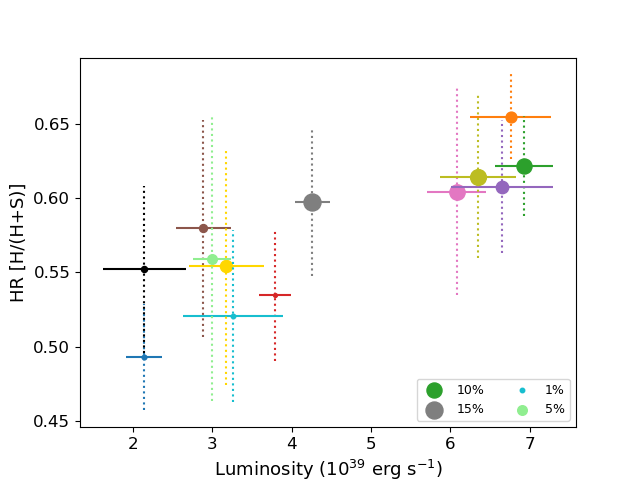}
    \caption{Average HR (from the 1.2--10/0.3--10 keV energy bands) versus luminosity in linear space. The size of the markers correspond to the different values of the fractional variability (reported in Table \ref{Fvar_mean}).}
    \label{Fvar_grafico}
\end{figure}

As shown in Fig. \ref{Fvar_grafico}, the variability of the source considerably increases in the observations with a higher flux, which is in agreement with the RMS-flux variability of accretion discs. \cite{Sutton_2013} suggest that the increased flux variability observed at energies above 1 keV in ULXs with hard spectra can be attributed to the obscuration of the hard central emission when observed through the clumpy edge of the outflowing wind and from the photosphere of the super-Eddington disc.

\cite{Sathyaprakash_2019} have found evidence of pulsations during observations Obs.ID:080399401 and Obs.ID:0803990601. This corresponds to the low-flux end of the new campaign, which suggests that the bright variable continuum is strongly affected by the inner accretion flow. In other words, at higher accretion rates the disc flux may significantly exceed the flux of the accretion column, thereby decreasing the pulse fraction. \cite{Walton_2018b} also suggest this for the ensemble of ULXs observed by {\it NuSTAR}. In addition, this would imply a low-to-mild magnetic field (B $\lesssim 10^{12}$ G, see e.g. \citealt{King_2020}), because the magnetospheric radius ($R_m$) is likely smaller than the spherisation radius ($R_{sph}$) for this object, given the presence of a bubble nebula (e.g. \citealt{Pakull_2002}) that is thought to be inflated by the disc wind.

\begin{table}[]
\centering
\caption{Fractional variability measured using the average value of the EPIC-pn light curve segments for each observation with an exposure time of 40 ks and 1000 s bins.}  
\begin{tabular}{lll}
\toprule
\toprule
Obs.ID & F$_{var}$ mean ($\%$) & HR mean \\
\midrule
0405090101  & 6.06 $\pm$ 0.02 & 0.65 $\pm$ 0.03 \\
0693850501  & 11.76 $\pm$ 0.07 & 0.62 $\pm$ 0.03 \\
0693851201  & 0.748 $\pm$ 0.008 & 0.53 $\pm$ 0.04 \\
0742590301  & 7.93 $\pm$ 0.07 & 0.61 $\pm$  0.05 \\    
0742490101  & 3.26 $\pm$ 0.04 & 0.58 $\pm$ 0.07 \\   
0764770101  & 1.154 $\pm$ 0.002 & 0.49 $\pm$  0.04 \\
0782310101  & 12.17 $\pm$ 0.09 & 0.60 $\pm$  0.07\\
0803990101  & 14.99 $\pm$ 0.04 & 0.60 $\pm$  0.05 \\
0803990201  & 12.93 $\pm$ 0.09 & 0.61 $\pm$  0.05 \\
0803990301  & 1.04 $\pm$ 0.04 & 0.52 $\pm$  0.06 \\
0803990401  & 1.91 $\pm$ 0.04 & 0.55 $\pm$  0.06 \\
0803990501  & 7.56 $\pm$ 0.16 & 0.55 $\pm$  0.08 \\
0803990601  & 4.79 $\pm$ 0.10 & 0.6 $\pm$  0.1 \\
\bottomrule
\end{tabular}
\label{Fvar_mean}
\end{table}

\section{Discussion}\label{DISCUSSION}

We have performed a detailed spectral analysis of the ultraluminous X-ray source NGC 1313 X-2, focussing on the \textit{XMM–Newton} observations, with the aim of understanding the long-term behaviour of the spectral components.
Given that the compact object is now known to be a NS, the accretion rate must be highly super-Eddington, and its spectral and temporal properties are expected to diverge from the case of sub-Eddington thin accretion discs.

As \cite{Vierdayanti_2010} showed, the evolution of NGC 1313 X-2 appears similar to the archetypal ULX Holmberg IX X-1, characterised by strong spectral variability (see also \citealt{Luangtip_2016, Pintore_2014, Walton_2014}). As reported by \cite{Walton_2017}, the spectra of Holmberg IX X-1 are well fit by two thermal blackbody components, which dominate the emission below 10 keV, plus a powerlaw tail which dominates above 10 keV.

As shown in Fig. \ref{grafico_confronto}, the spectrum is reasonably well described by a combination of two DISKBB plus a CUTOFFPL components. This model approximates super-Eddington accretion onto a NS, characterised by a standard outer disc, a thick inner disc region, and strong optically thick winds, which may contribute to the cooler component. In particular, the $\sim$ 1--10 keV band is mainly dominated by the hotter component, which describes the emission from the inner region, and the CUTOFFPL component representing the accretion column and the boundary layer near the NS.
Instead, the emission from the outer disc or from the wind is responsible for the cooler blackbody component with a temperature around 0.3 keV. Wind is expected to be launched from the upper regions of the inner disc at accretion rates comparable to or higher than the Eddington limit. This scenario is supported by the presence of strong and narrow residuals around 1 keV that have been resolved in similar sources with the aid of high-resolution X-ray spectra \citep{Pinto_2016}. 

The modelling of the spectra with high statistics shows that the column density, N$_H$, is broadly consistent with $\sim2\times10^{21}$ cm$^{-2}$ (see Table \ref{table_bestfit}). Fixing the N$_H$ (see Appendix \ref{appendix_nHfr}) does not strongly influence the broadband continuum fits, so we preferred to keep N$_H$ free.

With the intention of improving the analysis, we tested several spectral models, as described in Section \ref{Alternative models}. When the accretion rate is high, the structure of the disc is expected to deviate considerably from the standard Shakura-Sunayev thin disc. For this reason, as reported by \cite{Bachetti_2013}, we replaced the hotter DISKBB with a DISKPBB component (we tested the DISKBB+DISKPBB model). Considering only the DISKPBB component, we have obtained similar results. Comparing the same observations analysed in Bachetti et al. (Obs.ID1:0693850501 and Obs.ID2:0693851201), we found the values of $p$ to be consistent within $2\,\sigma$ (this work: p$_{\mathrm{ID}}1$=0.62$\pm$0.03 and p$_{\mathrm{ID}2}$=0.62${+0.18\atop-0.07}$; Bachetti et al.: p$_{\mathrm{ID}1}$=0.58$\pm$0.01 and p$_{\mathrm{ID}2}$=0.500${+0.006\atop*}$). We also found consistent temperatures (this work: T$_{in1}$=1.52${+0.06\atop-0.04}$ and T$_{in2}$=1.15${+0.06\atop-0.07}$; Bachetti et al.: T$_{in1}$=1.56$\pm$0.06 and T$_{in2}$=1.27$\pm$0.05). In general, for several observations, the $p$ values are compatible with an accretion regime significantly affected by advection and they are sometimes close to the slim disk regime, which strongly argues in favour of super-Eddington accretion.

In the future we will use principal component analysis (PCA) to decompose the spectrum in a model-independent way (e.g. \citealt{Pinto_2020b}).
Moreover, we will search for winds with both PCA and physical models and then we will study how they vary with the spectral shape. 

\subsection{Luminosity-temperature plane}

\begin{figure}[h!]
\centering
    \includegraphics[width=9cm]{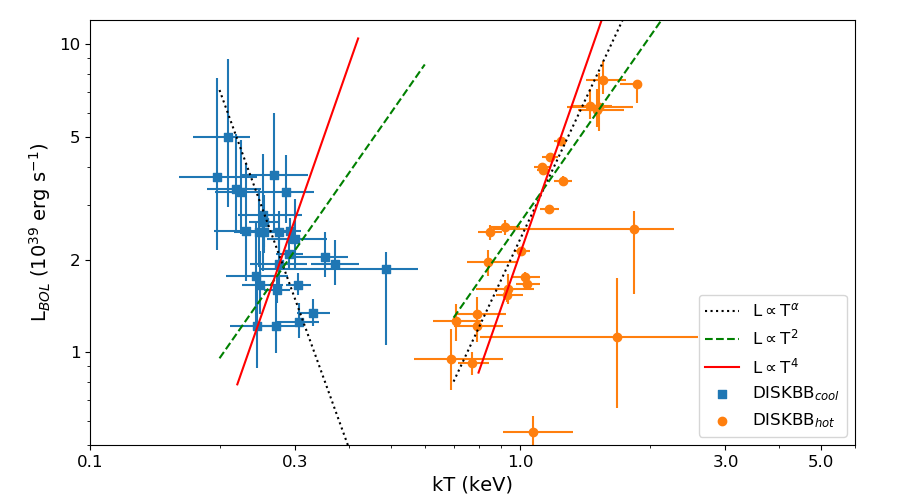}
    \caption{0.001–10 keV (i.e. bolometric) luminosity versus temperature for both the cool DISKBB (blue points) and hot DISKBB (orange points) components with free column density, N$_H$ (model: DISKBB+DISKBB+CUTOFFPL).}
    \label{LT1_Cutoffpl}
\end{figure}

In this section we discuss the temporal behaviour of the two thermal components, investigating how they evolve in the luminosity-temperature plane. We calculated the unabsorbed luminosities (i.e. corrected for interstellar absorption by setting the column density N$_H$=0) for each of the thermal components individually, over the broad band from 0.001–10 keV, which provides their bolometric values. The comparison between the temperature and the luminosity of each component in the case of free column density is shown in Fig. \ref{LT1_Cutoffpl}. 
We show the L--T trends for both the cool (blue) and hot (orange) thermal components in the same figure because their separation in temperature is sufficiently high to avoid confusion.

It is very useful to compare the observed trends with the L--T$^{\alpha}$ relationships as expected from theoretical scenarios such as a sub-Eddington thin disc of Shakura-Sunayev (L\,$\propto$\,T$^4$ for constant emitting area, see \citealt{Shakura_1973}) or an advection-dominated disc (L\,$\propto$\,T$^2$, see \citealt{Watarai_2000}). These laws are shown as red-solid and green-dashed lines in Fig. \ref{LT1_Cutoffpl}, respectively.

The L--T trend observed for the hot disk-blackbody component is somewhere in between the two trends above.\ We do indeed obtain $\alpha_H=3.0\pm0.35$ by fitting the L--T group of the hot component, which also argues in favour of super-Eddington accretion with a thicker disc. 

The trend followed by the cool component is instead negative ($\alpha_C=-3.9 \pm 1.0$) and differs form the one expected from a sub-Eddington thin disc of Shakura-Sunayev (L$\propto$T$^4$) or an advection-dominated disc (L$\propto$T$^2$). Our results are in agreement with the interpretation of \cite{Qiu_2021}, suggesting that the soft emission originates from the photosphere of the optically thick wind, driven by supercritical accretion.
In this scenario, the measured blackbody luminosities of the PULXs are often higher than the Eddington limit of NSs, assuming spherical accretion (i.e. L$_{\mathrm{EDD,NS}}$ = (1-3) $\times$ 10$^{38}$ erg s$^{-1}$). Moreover, we may expect an inversion of the L--T relationship at accretion rates much higher than the Eddington limit due to the expansion of the photosphere, which is marked by an increase in the size of the emitting region of the soft component and a decrease in temperature.
This effect can be interpreted as the fact that the spherisation radius, where the strong winds start to be launched, would increase with the accretion rate, yielding lower temperatures at a higher luminosity (see, e.g. \citealt{Poutanen_2007}.)

Although the nature of the compact object is unknown, the hotter disc component of NGC 1313 X-1 also exhibits a positive relation for the luminosity-temperature trend, consistent with the theoretical scenarios. \cite{Walton_2020} did not seem to find a strong anti-correlation between L and T for the low-T component, but this might be due to the low number of spectra (seven) along with the more complex DISKBB+SIMPL*DISKPBB and DISKPBB+DISKBB+CUTOFFPL model.

\begin{figure}[h!]
\centering
    \includegraphics[width=9cm]{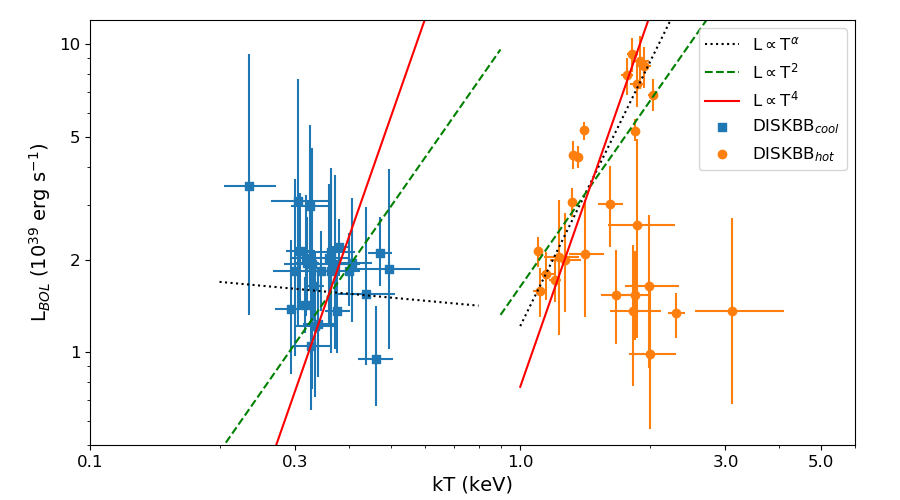}
    \caption{0.001–10 keV (i.e. bolometric) luminosity versus temperature for both the cool DISKBB (blue points) and hot DISKBB (orange points) components with free column density, N$_H$ (model: DISKBB+DISKBB).}
    \label{LT1}
\end{figure}

In order to check the presence of the third component, we tested the L--T trends with the simpler DISKBB+DISKBB model. As can be seen from Fig.\ref{LT1}, the L--T trend observed for the hot disk-blackbody component shows a much more chaotic behaviour. Observations with better statistics (higher fluxes and longer durations) still sit somewhere in between the two theoretical cases outlined above, but there are considerable deviations for a cluster of points where the hotter component is characterised by high temperatures and low luminosities. Indeed, we obtain $\alpha_H=2.8\pm0.6$ by fitting the L--T group of the hot component, which also argues in favour of super-Eddington accretion with a thicker disc. The trend followed by the cool component is instead much flatter ($\alpha_C=-0.13 \pm 0.36$) than the one expected from the theoretical scenarios.
We conclude that the introduction of a third (CUTOFFPL) component significantly improves the spectral fits of some observations, which is confirmed by the detection of more regular trends in Fig. \ref{LT1_Cutoffpl} and from their lower reduced $\chi^2$.

The presence of a correlation or anti-correlation between these points (luminosity and temperature for the hot and cool components) for the baseline model was also verified by the computation of the correlation coefficients of Pearsons and Spearmans (see Table \ref{pear_spear}). They were calculated using the \textsc{PYTHON} routines \emph{scipy.stats.pearsonr} and \emph{scipy.stats.spearmanr}. The result for the cooler component indicates a negative correlation and a positive correlation for the hotter component. The coefficients obtained are not very high because, as opposed to the least squares method, they do not take the error associated with the luminosity and temperature  into account.

\begin{table}[h!]
\centering 
\caption{Pearson and Spearman correlation coefficients calculated for the trends between luminosity and temperature for both the cool and the hot DISKBB components (assuming the baseline model, i.e. DISKBB+DISKBB+CUTOFFPL).} 
\begin{tabular}{lll}
\toprule
\toprule
Parameters & Pearson & Spearman \\
\midrule
L$_{hot}$--T$_{hot}$ & 0.69 & 0.69 \\
L$_{cool}$--T$_{cool}$ & -0.44 & -0.47 \\
\bottomrule
\end{tabular}
\label{pear_spear}
\end{table}

The behaviour of the soft component has also been recently examined by \cite{Gurpide_2021}. As they report, the significantly softer spectra of NGC 1313 X-2 can be explained by the scenario consistent with the wind structure responsible for highly anisotropic emission, given the wide HR variability the source spans. However, they do not report a clear correlation between the temperature and luminosity. Although their model is a double DISKBB model and thus is conceptually similar to ours, they adopted a different model for the high-energy component, using a SIMPL powerlaw continuum instead. As previously discussed, this may exacerbate parameter degeneracies when only soft X-ray data (i.e. <10 keV) are available, and this may explain the different conclusion found here. Moreover, for some observations they have used only pn spectra, resulting in a lower overall signal-to-noise ratio (S/N) and thus larger uncertainties. As we also suggest, they hypothesise that the observed residuals at soft energies around 1 keV can be produced by the presence of strong outflows.
\cite{Kajava_2009} also studied the (L,T) relation of the soft component, although much less data were available at that time (i.e. prior to our new 2017 campaign). They found that, taken together, the soft component followed a trend L$_{soft}$ $\propto$ T$^{-3.5}$ for a sample of ULXs that included NGC 1313 X-2, although the behaviour of individual sources was not considered in detail. In addition, \cite{Soria_2007} showed an anti-correlation between the disc luminosity and temperature, supporting the non-standard outer disc. As \cite{Soria_2007}  showed, if we assume that the boundary between the outer and inner disc scales as $\dot{m}$, the negative slope of the L--T relation depends on the T(R) function. For T $\sim$ R$^{-0.5}$, it is expected that the L $\sim$ T$^{-4}$, which is consistent with our result (L $\sim$ T$^{-3.9}$). Therefore, we suggest that the outer disc is affected by the wind.

About 10\% of the available data have also been analysed by \cite{Pintore_2012}. They adopted a different model where the hot DISKBB is replaced with the much broader Comptonisation component, COMPTT. They also tied the temperature of the seed photons to that of the cool blackbody. For the broad X-ray spectra of ULXs, this has the effect of lowering the contribution in terms of flux from the cooler blackbody component. They reported a weak correlation between the luminosity of the soft component with the inner temperature, L$_{disc} \propto$ T$_{disc}^{1.2\pm0.3}$.

\subsection{Radius-luminosity relation}

In order to test the consistency with the luminosity-temperature trends, we also estimated the mean inner radii for the DISKBB component for each of the two groups, using the normalisation factor with the formula $R_{in} = \sqrt{(norm*D^2_{10})/cos\,i}$, where $D_{10}$ is the distance to the source in units of 10 kpc and $i$ is the inclination of the disc. The mean inner radii are R$_{in,1}$ $\sim$ 1722 (cos\,$i$)$^{-1/2}$ km and R$_{in,2}$ $\sim$ 146 (cos\,$i$)$^{-1/2}$ km (where subscripts 1 and 2 refer to the lower and higher temperature tracks, respectively), which correspond to $\sim$ 414 R$_{S}$ and $\sim$ 35 R$_{S}$ (where $R_{S}= 2GM/c^2$ is the Schwarzshild radius, assuming M\,=\,1.4\,M$_\odot$).

\begin{figure}
    \centering
    \includegraphics[width=9cm]{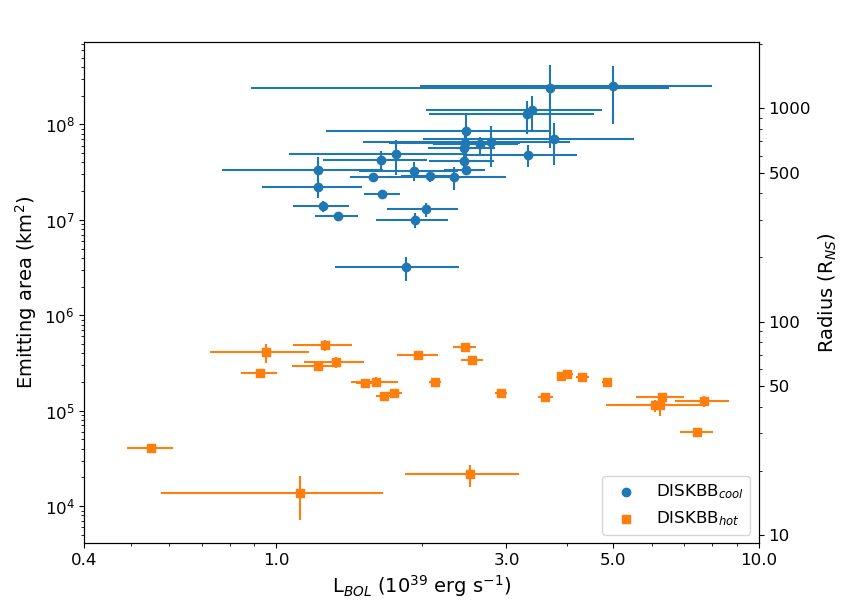}
    \caption{Left Y-axis: Emitting area versus luminosity for both the cool and hot DISKBB components (blue and orange points, respectively). Right Y-axis: Radius, in units of typical NS radius of 10 km, versus luminosity for both the cool and hot DISKBB components (blue and orange points, respectively).}
    \label{area_lum}
\end{figure}

Using the relation $L=A \sigma T^{4}$, we can also determine the emitting areas for each DISKBB component, as shown in Fig. \ref{area_lum}. We note that, for the hot component, the size of the emitting area of the accretion disc is broadly constant (in units of $R_{NS}$\,=10\,km, $\Delta R \sim 57\,R_{NS}$) for several observations. Instead, the emitting area for the cool component varies from R$_{min}$\,$\sim$\,180\,R$_{NS}$ to R$_{max}$\,$\sim$\,1600\,R$_{NS}$. This would agree with the expectations from a local increase in the accretion rate \citep{Poutanen_2007}.

Using the formalism outlined in \cite{King_2020} and assuming 20\,<\,$\dot{m}$\,<\,25 (which is reasonable for the observed luminosity), we can also estimate the spherisation radius R$_{sph}$ $\sim$ 70--87 R$_{S}$. We found that the inner radius for the cool component is larger than the spherisation radius, that is R$_{in,1}$ > R$_{sph}$, which suggests that the cool component takes both the emission from the cool disc and the contribution from the wind in
account. The inner radius of the hot DISKBB is instead smaller than the spherisation radius (R$_{in,2}$ < R$_{sph}$), indicating that this component is likely reproducing the super-Eddington inner accretion flow within R$_{sph}$.

Our approach has made use of phenomenological models in order to describe the long-term spectral evolution of the source.\ Future work will benefit from adopting physically motivated models which account for the thick nature of super-Eddington discs and Compton scattering through the disc atmosphere.

\section{Conclusions}\label{CONCLUSIONS}

In this paper we have presented the analysis of the X-ray emission spectrum of NGC 1313 X-2, a pulsating ultraluminous X-ray source, using all the available observations performed by \emph{XMM-Newton} between 2000 and 2017.
With the aim of characterising the spectral shape and its long-term variability, we have tested various spectral models for all observations. Since pulsations have been detected, we know that the compact object is a NS. In order to reach high luminosities (up to 10$^{40}$ erg s$^{-1}$ in some observations), the compact object must be in a super-Eddington accretion regime.

The spectral model that provides the best description of the {\it XMM-Newton}/EPIC data consists of two multi-colour disk blackbody components plus an exponential cutoff powerlaw, as in \cite{Walton_2018b}. Similar to previous works, we find that the hotter of the two thermal components dominates the 1--10 keV band, and the cooler one is more relevant in the energy band below 1 keV.
Furthermore, we found that the inner radius of the hot component is smaller than the spherisation radius, while the characteristic radii associated with the cooler component are all larger than the spherisation radius. In the framework of super Eddington accretion, this is strong evidence that the hotter component describes the dominant emission from the innermost region of the disc. Instead, the cooler component accounts for the emission from a larger region of the disc, outside the spherisation radius and the characteristic wind launching radius. Finally, we argued that the cutoff powerlaw component comes from the accretion columns onto the NS.

Most spectra show narrow residuals around $\sim$1 keV, which are likely associated with atomic emission and absorption lines produced by powerful outflows, as they are qualitatively similar to those seen in other ULXs that have been resolved and identified with the aid of dedicated high-resolution observations (see e.g. \citealt{Pinto_2020b}). The luminosity and the temperature of each emission component evolves along L--T trends that deviate from those expected for some typical regimes such as the sub-Eddington thin disc of Shakura-Sunayev (L\,$\propto$\,T$^4$, for constant emitting area) or the advection-dominated disc (L\,$\propto$\,T$^2$). In particular, we obtain power indexes of $\alpha_H=3.0\pm0.35$ and $\alpha_C=-3.9 \pm 1.0$ for the hotter and the cooler components, respectively. This implies a super-Eddington accretion regime and suggests a geometry closer to the funnel than a thin disk for the inner regions due to intense radiation pressure. Optically thick winds are also likely responsible for the scatter seen in Fig. \ref{LT1_Cutoffpl}.

 This behaviour is qualitatively similar to the broadband spectral evolution seen in NGC 1313 X-1 and Holmberg IX X-1. The similar spectral evolution between these sources, where the flux above 1 keV increases significantly at higher luminosity, may indicate a common structure and evolution among archetypal ULXs.

\begin{acknowledgements}
The authors would like to thank the anonymous referee, who provided useful suggestions for improving the final manuscript.
This work is based on observations obtained with {\it XMM-Newton}, an ESA science mission funded by ESA Member States and USA (NASA). DJW acknowledges support from STFC in the form of an Ernest Rutherford fellowship (ST/N004027/1). TPR gratefully acknowledges support from the Science and Technology Facilities Council (STFC) as part of the consolidated grant award ST/000244/1
\end{acknowledgements}

\bibliography{Bibliografia}
\bibliographystyle{aa}

\begin{appendix}
\section{Table best-fit parameters}\label{appendix_bestfit}
Table \ref{table_bestfit} reports the results of the spectral fits with the DISKBB+DISKBB+CUTOFFPL model for all {\it XMM-Newton} observations of X-2. 

\begin{landscape}
\begin{table}[h!]
\caption{Best fitting spectral parameters of NGC1313 X-2 in different observations obtained with the absorbed DISKBB+DISKBB+CUTOFFPL model.}
\begin{tabular}{llllllllll}
\toprule
\toprule
ObsID & N$_H$ & T$_1$ & norm$_1$ & L$_1$${(*)\atop}$ & T$_2$ & norm$_2$ & L$_2$${(*)\atop}$ & norm3 & $\chi^2$/dof \\
& (10$^{22}$ $\mathrm{cm^{-2}}$) & ($\mathrm{keV}$) & & (10$^{39}$ $\mathrm{erg/s}$) & ($\mathrm{keV}$) & & (10$^{39}$ $\mathrm{erg/s}$) & (10$^{-5}$) & \\
\midrule
0106860101 & 0.286${+0.06 \atop -0.04}$ & 0.219 $\pm$ 0.03 & 30${+57 \atop -16}$ & 3.4${+1.7 \atop -1.0}$ & 0.71${+0.10 \atop -0.08}$ & 0.11${+0.08 \atop -0.05}$ & 1.26${+0.18\atop -0.17}$ & 4.0 $\pm$ 0.3 & 137.31/144 \\ 
0150280101 & 0.10${+0.05 \atop -0.03}$ & 0.5${+1.8 \atop -0.5}$ & 0.13${+923 \atop -0.13}$ & 5.75${+24. \atop -5.}$ & 1.4${+8 \atop -1.3}$ & 0.03${+0.03 \atop -0.01}$ & 5.2${+1.1\atop -1.}$ & 1.0${+6.0 \atop -1.0}$ & 64.91/63 \\   
0150280301 & 0.19 $\pm$ 0.02 & 0.37 $\pm$ 0.05 & 2.3${+1.4 \atop -0.9}$ & 1.9${+0.4 \atop -0.3}$ & 1.871${+0.04 \atop -0.16}$ & 0.014${+0.004 \atop -0.002}$ & 7.46${+0.16\atop -1.}$ & 0.02${+3 \atop -0.02}$ & 161.42/191 \\   
0150280401 & 0.27${+0.06 \atop -0.05}$ & 0.27${+0.05 \atop -0.04}$ & 15${+33 \atop -10}$ & 3.8${+2. \atop -1.3}$ & 1.5${+0.3 \atop -0.2}$ & 0.027${+0.017 \atop -0.010}$ & 6.3${+1.8\atop -1.0}$ & 9${+5\atop -7}$ & 165.28/158 \\    
0150280501 & 0.13${+0.03 \atop -0.02}$ & 0.49${+0.09 \atop -0.2}$ & 0.005${+0.10 \atop -0.003}$ & 1.86${+0.3 \atop -0.8}$ & 1.8${+0.4 \atop -1.0}$ & 0.73${+41 \atop -0.35}$ & 2.5${+0.4\atop -1.}$ & 0.01${+6.0 \atop -0.01}$ & 86.77/87 \\   
0150280601 & 0.26${+0.06 \atop -0.05}$ & 0.23${+0.06 \atop -0.04}$ & 21${+48 \atop -15}$ & 2.47${+1.6 \atop -0.8}$ & 0.84${+0.14 \atop -0.09}$ & 0.09${+0.06 \atop -0.04}$ & 1.97${+0.18\atop -0.2}$ & 3.1${+0.6 \atop -0.7}$ & 151.03/122 \\    
0150281101 & 0.32${+0.09 \atop -0.05}$ & 0.2${+0.03 \atop -0.04}$ & 0.05${+0.05 \atop -0.018}$ & 5.${+4. \atop -2.}$ & 0.94 ${+0.14 \atop -0.15}$ & 50${+166 \atop -29}$ & 1.6$\pm$0.18 & 3.0${+0.9 \atop -1.0}$ & 62.86/82 \\    
0205230301 & 0.27 $\pm$ 0.03 & 0.29${+0.05 \atop -0.03}$ & 1.56${+0.2 \atop -0.14}$ & 3.32${+1.1 \atop -0.7}$ & 0.029${+0.009 \atop -0.008}$ & 11${+12 \atop -6}$ & 7.7${+1.2\atop -0.8}$ & 7${+3 \atop -5}$ & 212.25/210 \\   
0205230401 & 0.31${+0.08 \atop -0.07}$ & 0.20${+0.05 \atop -0.04}$ & 0.09${+0.17 \atop -0.07}$ & 3.7${+4. \atop -1.6}$ & 0.69${+0.2 \atop -0.12}$ & 55${+158 \atop -41}$ & 1.0$\pm$0.2 & 3.5${+0.5 \atop -0.6}$ & 82.88/90 \\     
0205230501 & 0.29${+0.05 \atop -0.04}$ & 0.22 $\pm$ 0.03 & 0.07${+0.06 \atop -0.03}$ & 3.3${+1.6 \atop -0.9}$ & 0.8${+0.12 \atop -0.09}$  & 28${+44 \atop -15}$ & 1.22$\pm$0.14 & 3.2$ \pm$ 0.4 & 130.98/128 \\     
0205230601 & 0.26 $\pm$ 0.05 & 0.25${+0.06 \atop -0.03}$ & 15${+24 \atop -11}$ & 2.8${+1.6 \atop -1.0}$ & 1.51${+0.2 \atop -0.15}$ & 0.026${+0.010 \atop -0.009}$ & 6.1${+1.1\atop -0.7}$ & 9${+3 \atop -4}$ & 178.94/213 \\     
0301860101 & 0.25 $\pm$ 0.03 & 0.30${+0.06 \atop -0.04}$ & 6${+9 \atop -4}$ & 2.3${+0.8 \atop -0.5}$ & 1.45${+0.19 \atop -0.14}$ & 0.03${+0.012 \atop -0.009}$ & 6.3${+0.9\atop -0.6}$ & 7${+3 \atop -4}$ & 210.74/218 \\    
0405090101 & 0.240${+0.016 \atop -0.015}$ & 0.275${+0.02 \atop -0.019}$ & 9${+5 \atop -3}$ & 2.3${+0.8 \atop -0.5}$ & 1.17${+0.06 \atop -0.05}$ & 0.051 $\pm$ 0.008 & 4.31${+0.14\atop -0.13}$ & 12.2 $\pm$ 0.7  & 350.97/300 \\    
0693850501 & 0.235${+0.015 \atop -0.014}$ & 0.290${+0.02 \atop -0.019}$ & 0.046${+0.006 \atop -0.005}$ & 2.1${+0.3 \atop -0.2}$ & 1.24 $\pm$ 0.04 & 6${+3 \atop -2}$ & 4.85${+0.11\atop -0.10}$ & 2.4${+0.5 \atop -0.6}$ & 297.79/284 \\    
0693851201 & 0.218 $\pm$ 0.013 & 0.31 $\pm$ 0.02 & 0.046${+0.009 \atop -0.008}$ & 1.66${+0.16 \atop -0.13}$ & 1.01${+0.05 \atop -0.04}$ & 4.3${+1.7 \atop -1.2}$ & 2.1$\pm$0.06 & 0.5${+0.2 \atop -0.3}$ & 265.81/235 \\    
0722650101 & 0.19${+0.05 \atop -0.03}$ & 0.31${+0.05 \atop -0.08}$ & 3.0${+13 \atop -1.5}$ & 1.25${+0.3 \atop -2e+12}$ & 1.0${+5 \atop -0.4}$ & 0.005${+2 \atop -0.005}$ & 0.17${+1.3\atop -2e+12}$ & 4.2${+0.5 \atop -4.2}$ & 63.42/72 \\    
0742590301 & 0.28${+0.03 \atop -0.02}$ & 0.3 $\pm$ 0.02 & 0.055${+0.010 \atop -0.009}$ & 2.6${+0.6 \atop -0.4}$ & 1.13 $\pm$ 0.05 & 15${+10 \atop -6}$ & 4.01$\pm$0.12 & 2.5 $\pm$ 0.6 & 281.44/231 \\     
0742490101 & 0.25 $\pm$ 0.03 & 0.3${+0.04 \atop -0.03}$ & 5${+5 \atop -3}$ & 1.2${+0.3 \atop -0.2}$ & 1.04${+0.08 \atop -0.06}$ & 0.032${+0.009 \atop -0.008}$ & 1.67$\pm$0.06 & 0.8 $\pm$ 0.3 & 226.31/197 \\   
0764770101 & 0.214${+0.018 \atop -0.017}$ & 0.272${+0.02 \atop -0.018}$ & 7${+3 \atop -2}$ & 1.59${+0.19 \atop -0.14}$ & 1.07${+0.3 \atop -0.16}$ & 0.009${+0.009 \atop -0.005}$ & 0.55${+0.07\atop -0.05}$ & 3.2${+0.3 \atop -0.4}$ & 173.59/227 \\   
0764770401 & 0.24${+0.05 \atop -0.04}$ & 0.24${+0.05 \atop -0.04}$ & 12${+23 \atop -7}$ & 1.8${+0.9 \atop -0.5}$ & 0.85${+0.06 \atop -0.05}$ & 0.11 $\pm$ 0.03 & 2.46${+0.13\atop -0.15}$ & 1.7 $\pm$ 0.4 & 178.34/167 \\    
0782310101 & 0.235${+0.013 \atop -0.012}$ & 0.292${+0.018 \atop -0.016}$ & 0.032${+0.006 \atop -0.005}$ & 2.5${+0.3 \atop -0.2}$ & 1.26 $\pm$ 0.06 & 7${+3 \atop -2}$ & 3.61${+0.14\atop -0.12}$ & 7.3${+0.6 \atop -0.7}$ & 339.58/290 \\    
0794580601 & 0.25${+0.05 \atop -0.04}$ & 0.25 $\pm$ 0.03 & 0.08${+0.03 \atop -0.02}$ & 2.5${+0.9 \atop -0.6}$ & 0.92 $\pm$ 0.07 & 15${+21 \atop -8}$ & 2.55${+0.13\atop -0.14}$ & 5.0 $\pm$ 0.6 & 193.82/187 \\    
0803990101 & 0.216${+0.014 \atop -0.013}$ & 0.33 $\pm$ 0.03 & 2.5${+1.3 \atop -0.8}$ & 11.34${+0.15 \atop -0.12}$ & 1.17${+0.06 \atop -0.05}$ & 0.035${+0.007 \atop -0.006}$ & 2.92$\pm$0.08 & 1.3${+0.4 \atop -0.5}$ & 229.80/249 \\    
0803990201 & 0.281${+0.019 \atop -0.018}$ & 0.254${+0.019 \atop -0.016}$ & 0.053 $\pm$ 0.007 & 2.5${+0.4 \atop -0.3}$ & 1.13 $\pm$  0.04 & 14${+7 \atop -5}$ & 3.90$\pm$0.09 & 2.8${+0.4 \atop -0.5}$ & 314.91/259 \\     
0803990301 & 0.225${+0.03 \atop -0.014}$ & 0.28${+0.05 \atop -0.04}$ & 7${+9 \atop -4}$ & 1.9${+0.6 \atop -0.3}$ & 0.80${+0.13 \atop -0.09}$ & 0.07${+0.06 \atop -0.04}$ & 1.33${+0.19\atop -0.19}$ & 3.4 $\pm$ 0.3 & 229.31/182 \\    
0803990401 & 0.27${+0.05 \atop -0.04}$ & 0.24${+0.04 \atop -0.03}$ & 0.056${+0.03 \atop -0.019}$ & 1.2${+0.6 \atop -0.3}$ & 0.8${+0.07 \atop -0.06}$ & 7${+11 \atop -4}$ & 0.93${+0.08\atop -0.08}$ & 2.3 $\pm$ 0.2 & 152.54/175 \\     
0803990701 & 0.20${+0.04 \atop -0.03}$ & 0.35${+0.05 \atop -0.07}$ & 0.004${+0.008 \atop -0.001}$ & 2.0${+0.4 \atop -0.3}$ & 1.7 $\pm$ 0.9 & 3.0${+6 \atop -1.3}$ & 1.1${+0.6\atop -0.5}$ & 1.5${+3 \atop -1.5}$ & 95.17/70 \\     
0803990501 & 0.27${+0.03 \atop -0.04}$ & 0.25${+0.03 \atop -0.02}$ & 10${+8 \atop -5}$ & 1.7${+0.5 \atop -0.3}$ & 0.93${+0.08 \atop -0.06}$ & 0.045 $\pm$ 0.013 & 1.53$\pm$0.07 & 0.9${+0.3 \atop -0.4}$ & 180.76/184 \\   
0803990601 & 0.210${+0.02 \atop -0.018}$ & 0.31${+ \atop }$ 0.03 & 3.18${+3 \atop -1.2}$ & 1.25${+0.19 \atop -0.14}$ & 1.03${+0.09 \atop -0.08}$ & 0.035${+0.013 \atop -0.010}$ & 1.75${+0.07\atop -0.08}$ & 0.7 $\pm$ 0.4 & 215.32/200 \\    
\bottomrule
\end{tabular}
\label{table_bestfit}
\begin{tablenotes}
   \item[*] \emph{Notes}: Parameter uncertainties were estimated at 68\%. ${(*)\atop}$Luminosity values (in units of 10$^{39}$ erg/s) are quoted for the unabsorbed model integrated over 0.3–-10 keV. 
  \end{tablenotes}
\end{table}
\end{landscape}

\begin{landscape}
\begin{figure*}[h!]
\centering
\scalebox{0.59}{\includegraphics{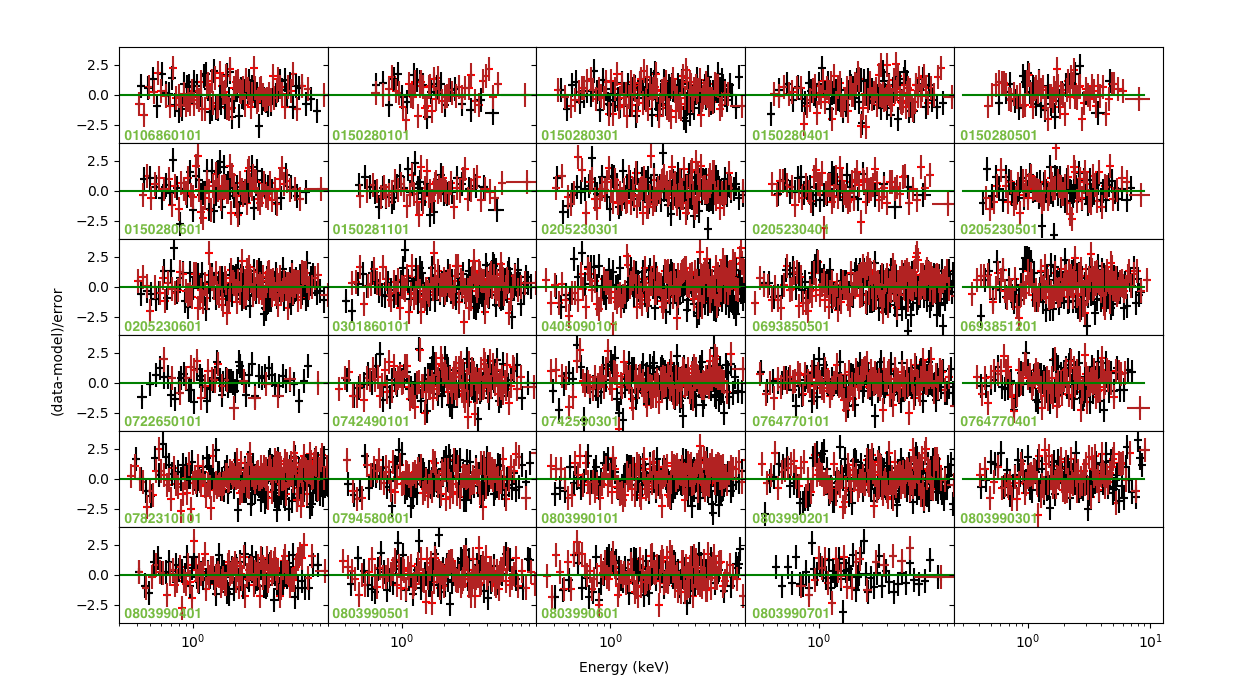}} 
\caption{Spectral residuals for the 29 spectra extracted from NGC 1313 X-2, calculated with respect to the DISKBB+DISKBB+CUTOFFPL model. Black and red points show data from the {\it XMM-Newton} EPIC-pn and EPIC-MOS detectors, respectively.}
\label{residui_tot}
\end{figure*}
\end{landscape}

\section{Table DISKBB+DISKBB parameters}\label{appendix_nHfree}
Table \ref{table_2diskbb} reports the results of the spectral fits with the DISKBB+DISKBB model for all {\it XMM-Newton} observations of X-2. Fig. \ref{residui_tot} shows the associated residuals for the best-fit model.

\begin{table*}[h!]
\centering
\caption{Best fitting spectral parameters of NGC1313 X-2 in different observations obtained with the absorbed DISKBB+DISKBB model.}
\begin{tabular}{lllllllll} 
\toprule
\toprule
ObsID & N$_H$ & T$_1$ & norm$_1$ & T$_2$ & norm$_2$ & $\chi^2$/dof \\
& (10$^{22}$ {cm$^{-2}$}) & ({keV}) & & ({keV}) & & \\
\midrule
0106860101 & 0.199${+0.018\atop -0.016}$ & 0.34${+0.02\atop -0.02}$ & 3.2${+1.4\atop -0.9}$ & 1.82${+0.17\atop -0.14}$ & 0.0031${+0.0012\atop -0.0009}$ & 149.14/145 \\   
0150280101 & 0.089${+0.2\atop -0.03}$ & 0.7${+6\atop -0.8}$ & 0.05${+101\atop -0.05}$ & 1.5${+1.5\atop -0.2}$ & 0.023${+0.018\atop -0.02}$ & 64.92/64 \\    
0150280301 & 0.19 $\pm$ 0.02 & 0.371 $\pm$ 0.05 & 2.3${+2\atop -1.1}$ & 1.87${+0.08\atop -0.07}$ & 0.014 $\pm$ 0.002 & 161.42/192 \\ 
0150280401 & 0.24 $\pm$ 0.04 & 0.3${+0.06\atop -0.04}$ & 8${+11\atop -5}$ & 1.90${+0.10\atop -0.09}$ & 0.016 $\pm$ 0.003 & 166.83/159 \\    
0150280501 & 0.50 $\pm$ 0.09 & 0.7${+0.8\atop -0.3}$ & 1.9${+0.4\atop -0.3}$ & 0.13${+0.03\atop -0.02}$ & 0.005${+0.004\atop -0.003}$ & 86.77/88 \\   
0150280601 & 0.19${+0.03\atop -0.02}$ & 0.363 $\pm$ 0.05 & 2.4${+2\atop -1.1}$ & 1.42${+0.15\atop -0.12}$ & 0.012${+0.006\atop -0.004}$ & 157.38/123 \\ 
0150281101 & 0.29 $\pm$ 0.06 & 0.23${+0.04\atop -0.03}$ & 26${+43\atop -16}$ & 1.27${+0.12\atop -0.10}$ & 0.017${+0.007\atop -0.006}$ & 66.69/83 \\ 
0205230301 & 0.25${+0.03\atop -0.02}$ & 0.33${+0.04\atop -0.03}$ & 6${+5\atop -3}$ & 1.82${+0.06\atop -0.05}$ & 0.020${+0.003\atop -0.002}$ & 213.96/211 \\ 
0205230401 & 0.20${+0.04\atop -0.03}$ & 0.30${+0.04\atop -0.03}$ & 5${+5\atop -3}$ & 1.8${+0.3\atop -0.2}$ & 0.0028${+0.0018\atop -0.0012}$ & 88.09/91 \\   
0205230501 & 0.213 $\pm$ 0.02 & 0.32 $\pm$ 0.03 & 4.4${+3\atop -1.6}$ & 1.67${+0.16\atop -0.13}$ & 0.0045${+0.0018\atop -0.0014}$ & 142.92/129 \\   
0205230601 & 0.21${+0.04\atop -0.03}$ & 0.33${+0.07\atop -0.05}$ & 4${+5\atop -2}$ & 1.94${+0.08\atop -0.07}$ & 0.014$\pm$ 0.002 & 181.97/214 \\    
0301860101 & 0.225 $\pm$ 0.02 & 0.36${+0.05\atop -0.04}$ & 2.7${+2\atop -1.3}$ & 1.77 $\pm$0.06 & 0.019 $\pm$ 0.003 & 213.29/219 \\ 
0405090101 & 0.182${+0.008\atop -0.007}$ & 0.47 $\pm$ 0.03 & 0.9${+0.3\atop -0.2}$ & 2.03 $\pm$ 0.06 & 0.0095${+0.0012\atop -0.0011}$ & 411.82/301 \\   
0693850501 & 0.213 $\pm$ 0.010 & 0.35 $\pm$ 0.02 & 2.8${+1.0\atop -0.7}$ & 1.41 $\pm$ 0.02 & 0.030 $\pm$ 0.002 & 310.06/285 \\  
0693851201 & 0.208${+0.011\atop -0.010}$ & 0.331${+0.019\atop -0.017}$ & 3.0${+0.9\atop -0.7}$ & 1.10${+0.03\atop -0.02}$ & 0.033$\pm$ 0.004 & 269.55/236 \\    
0722650101 & 0.175 $\pm$ 0.03 & 0.333${+0.04\atop -0.03}$ & 2.3${+1.7\atop -1.0}$ & 3.1${+1.0\atop -0.6}$ & 0.0004${+0.0004\atop -0.0002}$ & 63.79/73 \\    
0742590301 & 0.245${+0.018\atop -0.017}$ & 0.31${+0.03\atop -0.02}$ & 5.5${+3\atop -1.9}$ & 1.33 $\pm$ 0.03 & 0.0316${+0.0033\atop -0.0032}$ & 292.76/232 \\    
0742490101 & 0.218${+0.02\atop -0.018}$ & 0.33 $\pm$ 0.03 & 2.1${+1.4\atop -0.8}$ & 1.21 $\pm$ 0.04 & 0.018 $\pm$ 0.003 & 230.70/198 \\ 
0764770101 & 0.186$\pm$ 0.011 & 0.316$\pm$ 0.012 & 3.2${+0.7\atop -0.6}$ & 2.30${+0.11\atop -0.10}$ & 0.0012$\pm$ 0.0002 & 182.24/228 \\   
0764770401 & 0.167${+0.02\atop -0.017}$ & 0.44 $\pm$ 0.07 & 0.9${+0.8\atop -0.4}$ & 1.23${+0.14\atop -0.09}$ & 0.020${+0.011\atop -0.009}$ & 186.74/168 \\  
0782310101 & 0.201$\pm$ 0.008 & 0.380${+0.018\atop -0.017}$ & 2.3${+0.6\atop -0.4}$ & 1.85 $\pm$ 0.04 & 0.0103${+0.0009\atop -0.0008}$ & 380.93/291 \\  
0794580601 & 0.175${+0.018\atop -0.016}$ & 0.41${+0.05\atop -0.04}$ & 1.6${+1.0\atop -0.6}$ & 1.62${+0.12\atop -0.10}$ & 0.010 $\pm$0.003 & 213.02/188 \\  
0803990101 & 0.204${+0.011\atop -0.010}$ & 0.38 $\pm$ 0.03 & 1.5${+0.6\atop -0.4}$ & 1.32 $\pm$ 0.03 & 0.023${+0.003\atop -0.002}$ & 235.32/250 \\  
0803990201 & 0.242${+0.014\atop -0.013}$ & 0.319${+0.02\atop -0.019}$ & 4.2${+1.7\atop -1.2}$ & 1.36${+0.03\atop -0.02}$ & 0.028 $\pm$ 0.002 & 336.86/260 \\    
0803990301 & 0.176${+0.013\atop -0.012}$ & 0.40${+0.03\atop -0.02}$ & 1.6${+0.5\atop -0.4}$ & 1.84${+0.17\atop -0.14}$ & 0.0031${+0.0012\atop -0.0009}$ & 242.51/183 \\ 
0803990401 & 0.177${+0.016\atop -0.014}$ & 0.46 $\pm$ 0.04 & 0.46${+0.2\atop -0.14}$ & 2.0${+0.3\atop -0.2}$ & 0.0014${+0.0009\atop -0.0006}$ & 173.38/176 \\   
0803990701 & 0.19 $\pm$ 0.03 & 0.359 $\pm$ 0.04 & 2.7${+2\atop -1.1}$ & 2.0${+0.3\atop -0.2}$ & 0.0024${+0.0017\atop -0.0011}$ & 95.21/71 \\    
0803990501 & 0.23 $\pm$ 0.02 & 0.29 $\pm$ 0.03 & 4.1${+3\atop -1.6}$ & 1.11 $\pm$ 0.04 & 0.0230 $\pm$ 0.004 & 185.33/185 \\   
0803990601 & 0.198${+0.016\atop -0.015}$ & 0.34 $\pm$ 0.03 & 2.1${+1.1\atop -0.7}$ & 1.15${+0.05\atop -0.04}$ & 0.023 $\pm$ 0.004 & 217.35/201 \\   
\bottomrule
\end{tabular}
\label{table_2diskbb}
\begin{tablenotes}
   \item[*] \emph{Notes}: Parameter uncertainties were estimated at 68\%. 
\end{tablenotes}
\end{table*}

\begin{landscape}
\begin{figure*}[h!]
\centering
\scalebox{0.59}{\includegraphics{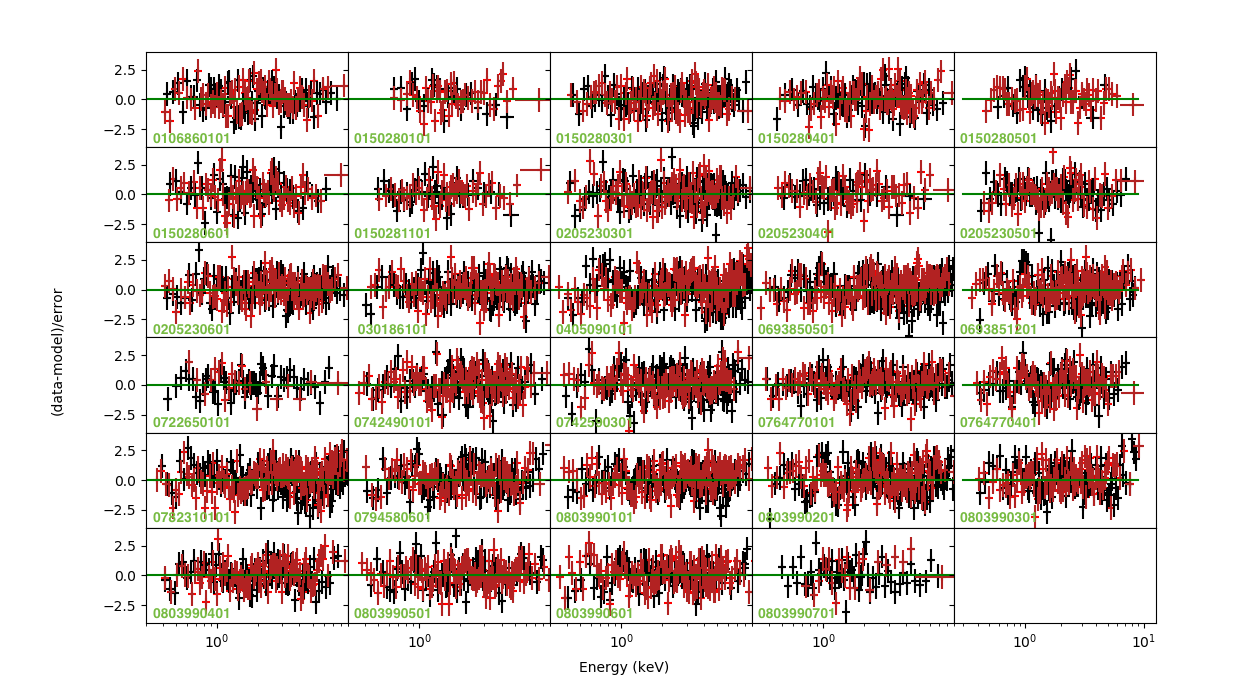}} 
\caption{Spectral residuals for the 29 spectra extracted from NGC 1313 X-2, calculated with respect to the DISKBB+DISKBB model. Black and red points show data from the {\it XMM-Newton} EPIC-pn and EPIC-MOS detectors, respectively.}
\label{residui_tot}
\end{figure*}
\end{landscape}

\section{Table of fractional variability, F$_{var}$}\label{appendix_Fvar}

\begin{table}[h!]
\caption{Fractional variability ($\%$) measured using 40 ks EPIC-pn segments of light curve for each observation, which is a common time baseline for comparing the RMS estimated.}

\begin{tabular}{lllll}
    \toprule
    \toprule
    Obs.ID &  1st segment (40ks) & 2nd segment (40/80ks) & 3rd segment (80/120ks) \\ 
    \midrule
        0106860101 &    -                &      -                & - \\
        0150280101 &    -                &      -                & - \\
        0150280301 &    -                &      -                & - \\
        0150280401 &    -                &      -                & - \\
        0150280501 &    -                &      -                & - \\
        0150280601 &    -                &      -                & - \\
        0150281101 &    -                &      -                & - \\
        0205230301 &    -                &      -                & - \\
        0205230401 &    -                &      -                & - \\
        0205230501 &    -                &      -                & - \\
        0205230601 &    -                &      -                & - \\
        0301860101 &    -                &      -                & - \\
        0405090101 & 10.98 $\pm$ 0.02  & 7.191 $\pm$ 0.012 & - \\
        0693850501 & 18.22 $\pm$ 0.04  & 17.05 $\pm$ 0.05  & - \\ 
        0693851201 & 2.245 $\pm$ 0.008 &        -                & - \\ 
        0722650101 &    -                &      -                & - \\
        0742590301 & 23.79 $\pm$ 0.07  &        -                & - \\
        0742490101 & 4.16  $\pm$ 0.04  & 5.616 $\pm$ 0.011 & - \\
        0764770101 & 3.463 $\pm$ 0.002 &        -                & - \\
        0764770401 &    -                &      -                & - \\
        0782310101 & 11.98 $\pm$ 0.04  & 24.53 $\pm$ 0.07        & - \\
        0794580601 &    -                &      -                & - \\
        0803990101 & 15.90 $\pm$ 0.02  & 9.333 $\pm$ 0.009 & 19.73 $\pm$ 0.03 \\ 
        0803990201 & 4.907 $\pm$ 0.003 & 21.97 $\pm$ 0.09  & 11.92 $\pm$ 0.02 \\ 
        0803990301 & 3.12 $\pm$ 0.04     &      -                & - \\
        0803990401 & 5.743 $\pm$ 0.004 &        -                & - \\
        0803990701 &    -                &      -                & - \\ 
        0803990501 & 18.87 $\pm$ 0.04  & 3.82 $\pm$ 0.15         & - \\
        0803990601 & 14.36 $\pm$ 0.10  &        -                & - \\
\bottomrule
\end{tabular}
\label{Fvar}
\end{table}

\clearpage
\newpage
\section{Table DISKBB+DISKBB parameters with fixed N$_H$}\label{appendix_nHfr}
Table \ref{table_bestfit_nHfr} reports the results of the spectral fits with the DISKBB+DISKBB model with fixed N$_H$ (N$_H$ = (0.194$\times$10$^{22}$) cm$^{-2}$) for all {\it XMM-Newton} observations of X-2.

\begin{table}[h!]
\centering
\caption{Best fitting spectral parameters of NGC1313 X-2 in different observations obtained with the absorbed DISKBB+DISKBB model with fixed N$_H$.}
\begin{tabular}{lllllll}
\toprule
\toprule
ObsID & N$_H$ & T$_1$ & norm$_1$ & T$_2$ & norm$_2$ & $\chi^2$/dof \\
& (10$^{22}$ {cm$^{-2}$}) & ({keV}) & & ({keV}) & &  \\
\midrule
0106860101 & 0.194 & 0.348 $\pm$ 0.011 & 2.9${+0.4\atop -0.3}$ & 1.85${+0.14\atop -0.12}$ & 0.0029${+0.0009\atop -0.0007}$ & 149.22/146 \\   
0150280101 & 0.194 & 0.22${+0.06\atop -0.04}$ & 14${+25\atop -9}$ & 1.38${+0.09\atop -0.08}$ & 0.037${+0.005\atop -0.009}$ & 65.24/65 \\    
0150280301 & 0.194 & 0.34${+0.03\atop -0.02}$ & 2.4${+0.7\atop -0.5}$ & 1.87${+0.07\atop -0.06}$ & 0.0141${+0.0020\atop -0.0019}$ & 161.42/193 \\   
0150280401 & 0.194 & 0.37${+0.04\atop -0.03}$ & 2.4${+0.9\atop -0.7}$ & 1.96${+0.10\atop -0.09}$ & 0.014 $\pm$ 0.003 & 168.43/160 \\    
0150280501 & 0.194 & 0.35 $\pm$ 0.03 & 3.5${+1.5\atop -1.0}$ & 1.53${+0.18\atop -0.14}$ & 0.0121${+0.006\atop -0.005}$ & 90.97/89 \\    
0150280601 & 0.194 & 0.351 $\pm$ 0.02 & 2.8${+0.7\atop -0.5}$ & 1.39${+0.11\atop -0.09}$ & 0.0126 $\pm$ 0.004 & 157.44/124 \\   
0150281101 & 0.194 & 0.30 $\pm$ 0.02 & 4.9${+1.6\atop -1.2}$ & 1.39${+0.14\atop -0.11}$ & 0.011${+0.005\atop -0.004}$ & 69.78/84 \\ 
0205230301 & 0.194 & 0.42 $\pm$ 0.03 & 1.4${+0.4\atop -0.3}$ & 1.89 $\pm$ 0.06 & 0.016 $\pm$ 0.002 & 220.15/212 \\  
0205230401 & 0.194 & 0.305${+0.016\atop -0.015}$ & 4.6${+1.1\atop -0.9}$ & 1.86${+0.2\atop -0.19}$ & 0.0027${+0.0014\atop -0.0010}$ & 88.11/92 \\   
0205230501 & 0.194 & 0.341${+0.013\atop -0.012}$ & 3.1${+0.5\atop -0.4}$ & 1.75${+0.14\atop -0.12}$ & 0.0037${+0.0013\atop -0.0010}$ & 143.70/130 \\    
0205230601 & 0.194 & 0.365 $\pm$ 0.03 & 2.0${+0.6\atop -0.5}$ & 1.97${+0.07\atop -0.06}$ & 0.0131${+0.0017\atop -0.0016}$ & 182.19/215 \\   
0301860101 & 0.194 & 0.44 $\pm$ 0.03 & 1.0${+0.3\atop -0.2}$ & 1.82 $\pm$ 0.06 & 0.016 $\pm$ 0.002 & 216.03/220 \\  
0405090101 & 0.194 & 0.432$\pm$ 0.012 & 1.38${+0.12 \atop -0.13}$ & 1.981 $\pm$ 0.04 & 0.0106$\pm$ 0.0008 & 414.11/302 \\   
0693850501 & 0.194 & 0.386$\pm$ 0.011 & 1.60${+0.16\atop -0.15}$ & 1.434${+0.019\atop -0.018}$ & 0.0273$\pm$ 0.0016 & 314.18/286 \\ 
0693851201 & 0.194 & 0.354$\pm$ 0.009 & 2.15${+0.2\atop -0.18}$ & 1.12 $\pm$ 0.02 & 0.030 $\pm$ 0.003 & 271.48/237 \\   
0722650101 & 0.194 & 0.315${+0.016\atop -0.015}$ & 3.1${+0.7\atop -0.6}$ & 2.9${+0.7\atop -0.4}$ & 0.0005${+0.0004\atop -0.0003}$ & 64.17/74 \\ 
0742590301 & 0.194 & 0.398${+0.017\atop -0.016}$ & 1.42${+0.2\atop -0.18}$ & 1.39 $\pm$ 0.03 & 0.025 $\pm$ 0.003 & 303.95/233 \\
0742490101 & 0.194 & 0.368${+0.017\atop -0.016}$ & 1.12${+0.19\atop -0.16}$ & 1.238 $\pm$ 0.04 & 0.016 $\pm$ 0.002 & 232.48/199 \\  
0764770101 & 0.194 & 0.308$\pm$ 0.005 & 3.7 $\pm$ 0.3 & 2.27 $\pm$ 0.09 & 0.0012$\pm$ 0.0002 & 182.73/229 \\    
0764770401 & 0.194 & 0.36$\pm$ 0.02 & 2.1${+0.5\atop -0.4}$ & 1.14${+0.06\atop -0.05}$ & 0.030${+0.007\atop -0.006}$ & 188.30/169 \\    
0782310101 & 0.194 & 0.393${+0.009\atop -0.008}$ & 1.95${+0.16\atop -0.14}$ & 1.870 $\pm$ 0.03 & 0.0099$\pm$ 0.0007 & 381.64/292 \\ 
0794580601 & 0.194 & 0.368$\pm$ 0.016 & 2.5 $\pm$ 0.4 & 1.55${+0.07\atop -0.06}$ & 0.013${+0.003\atop -0.002}$ & 214.05/189 \\  
0803990101 & 0.194 & 0.398$\pm$ 0.014 & 1.14${+0.14\atop -0.12}$ & 1.335 $\pm$ 0.03 & 0.021 $\pm$ 0.002 & 236.22/251 \\ 
0803990201 & 0.194 & 0.414${+0.015\atop -0.014}$ & 1.13${+0.13\atop -0.11}$ & 1.43 $\pm$ 0.03 & 0.0220${+0.0019\atop -0.0018}$ & 354.12/261 \\  
0803990301 & 0.194 & 0.370$\pm$ 0.011 & 2.3${+0.3\atop -0.2}$ & 1.73${+0.11\atop -0.10}$ & 0.0041${+0.0011\atop -0.0009}$ & 244.50/184 \\   
0803990401 & 0.194 & 0.370$\pm$ 0.011 & 0.422$\pm$ 0.018 & 0.69${+0.11\atop -0.09}$ & 1.85${+0.16\atop -0.13}$ & 174.49/177 \\  
0803990701 & 0.194 & 0.357${+0.019\atop -0.018}$ & 2.8${+0.6\atop -0.5}$ & 2.0${+0.3\atop -0.2}$ & 0.0025${+0.0014\atop -0.0010}$ & 95.22/72 \\ 
0803990501 & 0.194 & 0.344${+0.015\atop -0.014 }$ & 1.8${+0.3\atop -0.2}$ & 1.16 $\pm$ 0.04 & 0.018 $\pm$ 0.003 & 188.71/186 \\ 
0803990601 & 0.194 & 0.346$\pm$ 0.014 & 1.9${+0.3\atop -0.2}$ & 1.16${+0.04\atop -0.03}$ & 0.022 $\pm$ 0.003 & 217.43/202 \\    
\bottomrule
\end{tabular}
\label{table_bestfit_nHfr}
\begin{tablenotes}
  \item[*] \emph{Notes}: Parameter uncertainties were estimated at 68\%. 
\end{tablenotes}
\end{table}

\end{appendix}

\end{document}